\documentclass[a4paper,11pt,table]{article}

\pdfoutput=1 

\usepackage{jheppub} 

\usepackage[T1]{fontenc} 

\usepackage{xcolor}
\usepackage{draft} 
\usepackage{hyperref}
\usepackage{graphicx,color}
\usepackage{tcolorbox}
\usepackage{tikz-cd} 
\usepackage[all]{xy}
\usepackage{array,multirow}

\title{\boldmath Classification of  Argyres-Douglas theories from M5 branes}

\author[a]{Yifan Wang}
\author[b,c]{Dan Xie}

\affiliation[a]{Center for Theoretical Physics, Massachusetts Institute of Technology,Cambridge, 02139, USA}
\affiliation[b]{Center of Mathematical Sciences and Applications, Harvard University, Cambridge, 02138, USA}
\affiliation[c]{Jefferson Physical Laboratory, Harvard University, Cambridge, MA 02138, USA}

\emailAdd{yifanw@mit.edu}
\emailAdd{dxie@cmsa.fas.harvard.edu}

\abstract{We obtain a large class of new 4$d$ Argyres-Douglas theories by classifying 
 	irregular punctures for the 6$d$ $(2,0)$ superconformal theory of ADE type on a sphere. Along the way, we identify the connection between the Hitchin system and three-fold singularity descriptions of the same Argyres-Douglas theory. Other constructions such as taking degeneration limits of the
 	irregular puncture, adding an extra regular puncture, and introducing outer-automorphism twists are also discussed. Later we investigate various features of these theories including their Coulomb branch spectrum and central charges.
}

\begin{document} 
	\rightline{MIT-CTP/4711}
	\maketitle
	
	\flushbottom

\section{Introduction}\label{sect:I}
Four dimensional $\mathcal{N}=2$ superconformal field theories (SCFT) play an unusual role in the space of SCFTs across various dimensions:
(1). there are various interesting IR dynamics such as the existence of a moduli space of vacua with many interesting phase structures;  (2). often many 
aspects of the theories such as low energy effective action and BPS spectrum can be solved exactly.  
The investigation of these theories can teach us many lessons about generic features of quantum field theory such as confinement, renormalization group flow, etc. 

The space of $\mathcal{N}=2$ SCFTs has enlarged tremendously since the discovery of 
Seiberg-Witten (SW) solutions \cite{Seiberg:1994aj,Seiberg:1994rs}. One can engineer many Lagrangian theories from $\cN=2$ Super Yang-Mills (SYM) coupled to free hypermultiplets. More recently, it was found 
that strongly coupled matter systems such as $T_N$ theories can also be used to construct new SCFTs \cite{Gaiotto:2009we}, known as Class S theories, which greatly increases
the repertoire of known theories.  These theories have one distinguished feature that all BPS operators on the Coulomb branch\footnote{
	By BPS operators on the Coulomb branch we are referring to the Lorentz-scalar chiral primaries of $\cN=2$ superconformal algebra whose expectation values parametrize the Coulomb branch of the SCFT. These operators are annihilated by the four anti-chiral Poincar\'{e} supercharges and are invariant under $SU(2)_R$. In the rest of the paper, we will use ``chiral primary'' and ``BPS operator'' interchangeably to denote these operators. 
} have 
integer scaling dimensions. 

There is another type of $\mathcal{N}=2$ SCFTs called Argyres-Douglas (AD) theories \cite{Argyres:1995jj}. These
are {\it intrinsically} strongly coupled theories, the first instance of which 
was originally discovered at a certain point on the Coulomb branch of the pure $\cN=2$ $SU(3)$ theory.  Unlike familiar SCFTs,  such as $\cN=2$ $SU(2)$ SYM coupled to four fundamental 
flavors, the AD theory has some unusual properties: first the scaling dimensions of the Coulomb branch BPS operators are fractional, and second there are relevant 
operators in the spectrum. These special features make AD theories a particularly useful class of models from which we can study
generic features of conformal field theory, for example the RG flow between various fixed points \cite{Xie:2013jc}.  By looking at special points on the Coulomb branch of other gauge theories, 
new examples with an ADE classification has been found \cite{Eguchi:1996vu,Eguchi:1996ds}.  More recently, a further remarkable generalization 
of these theories which are called $(G, G')$ theories\footnote{
	Here the label means that the BPS quiver has the shape of the product of $G$ and $G'$ Dynkin diagram.} 
were engineered using type IIB string theory construction \cite{Shapere:1999xr,Cecotti:2010fi}. 

In \cite{Xie:2012hs},  a large class of new $\mathcal{N}=2$ AD theories has been constructed using $A$ type M5 brane construction (see \cite{Gaiotto:2009hg,Cecotti:2011rv,Bonelli:2011aa} for theories engineered using $A_1$ $(2,0)$ theory): one can engineer four dimensional 
SCFTs by putting 6$d$ $A_{n-1}$ (2,0)  theory on a punctured Riemann surface. To get a SCFT (within the construction of \cite{Xie:2012hs}), one must use a sphere with one irregular puncture, or 
a sphere with one irregular and one regular puncture. The classification of irregular punctures is very rich,  and, in particular one can also 
reproduce theories which are originally engineered using regular punctures on sphere \cite{Xie:2012hs}. 
These lessons suggest that the Argyres-Douglas theories constitute actually a much larger class of $\mathcal{N}=2$ theories than the usual Class S theories with 
integral Coulomb branch spectrum, and it is interesting to further enlarge the theory space.  

The main purpose of this paper is to generalize the $A$ type construction of \cite{Xie:2012hs} to other types of $(2,0)$ theory labeled by $J$. The major problem in the
M5 brane construction is to classify the irregular singularities \footnote{
	We use ``puncture'' and ``singularity'' interchangeably to denote the singular behavior of $\Phi$.}, which we find to take the following 
universal form: 
\begin{equation}
\Phi={T\over z^{2+k/b}}+\cdots.
\label{hp}
\end{equation}
Here $\Phi$ is the holomorphic part of the Higgs field which describes the transversal deformation of M5 branes of $J=A,D,E$ type. $T$ is a regular semi-simple 
element of $J$ and  $k>-b$ is an arbitrary integer. The allowed set of values for $b$ is given in Table~\ref{table:slopes1}  and the choice of $T$ in general depends on that of $b$. 

\begin{table}[!htb]
	\begin{center}
		\begin{tabular}{ |c|c| }
			\hline
			$J$ & allowed $b$  \\ \hline
			$A_{N-1}$& $N,~N-1$ \\ \hline
			$D_N$   & $2N-2~N$  \\     \hline
			$E_6$   & $12,~9,~8 $ \\     \hline
			$E_7$   & $18,~14$  \\     \hline
			$E_8$   & 30,~24,~20 \\     \hline
		\end{tabular}
	\end{center}
	\caption{Allowed denominators for irregular singularities.}
	\label{table:slopes1}
\end{table}
Given the structure of the singularity, one can obtain the SW solution by computing the spectral curve of the Hitchin system $det(x-\Phi)=0$. Moreover, we 
can directly map the corresponding $\cN=2$ AD theory to a 3-fold isolated singularity using the spectral curve (see Table~\ref{table:isolatedsingularitiesALEfib} and Figure~\ref{fig:ADsh}).\footnote{
	Here the matching to 3-fold singularities is restricted to $\cN=2$ AD theories defined by a single singularity of the form \eqref{hp} on $\mP^1$ (the Higgs field is smooth elsewhere) which requires $k\geq 1$.
  Note that for $b=h$ and $k=1$ the 3-fold hypersurface is nonsingular and the corresponding $4d$ theory is empty. }  
\begin{table}[!htb]
	\begin{center}
		\begin{tabular}{ |c|c|c| }
			\hline
			$J$& Singularity & $b$  \\ \hline
			$A_{N-1}$ &$x_1^2+x_2^2+x_3^N+z^k=0$&  $N$ \\ \hline
			$~$& $x_1^2+x_2^2+x_3^N+x_3 z^k=0$ & $N-1$\\ \hline
			
			$D_N$   & $x_1^2+x_2^{N-1}+x_2x_3^2+z^k=0$ & $2N-2$ \\     \hline
			$~$   &$x_1^2+x_2^{N-1}+x_2x_3^2+z^k x_3=0$& $N$ \\     \hline
			
			$E_6$  & $x_1^2+x_2^3+x_3^4+z^k=0$&12   \\     \hline
			$~$  & $x_1^2+x_2^3+x_3^4+z^k x_3=0$ &9   \\     \hline
			$~$  & $x_1^2+x_2^3+x_3^4+z^k x_2=0$  &8   \\     \hline
			
			$E_7$  & $x_1^2+x_2^3+x_2x_3^3+z^k=0$& 18   \\     \hline
			$~$  & $x_1^2+x_2^3+x_2x_3^3+z^kx_3=0$ &14    \\     \hline

			$E_8$   & $x_1^2+x_2^3+x_3^5+z^k=0$&30   \\     \hline
			$~$   & $x_1^2+x_2^3+x_3^5+z^k x_3=0$ & 24   \\     \hline
			$~$   & $x_1^2+x_2^3+x_3^5+z^k x_2=0$ & 20  \\     \hline
			
		\end{tabular}
	\end{center}
	\caption{3-fold singularities corresponding to our irregular punctures.}
	\label{table:isolatedsingularitiesALEfib}
\end{table}

Once the basic irregular punctures are identified, there are three variations which give rise to new theories:
\begin{itemize}
	\item We can take degeneration limits for some punctures, namely, the eigenvalues of the matrices in defining the irregular singularity can become degenerate.
	\item We can add another regular puncture, so as to obtain theories with non-Abelian flavor symmetry.
	\item If $J$ has a nontrivial outer-automorphism group, we can introduce twists on the punctures. This is only possible for a sphere with one irregular and one regular puncture. 
\end{itemize}
Using these constructions, we can find a lot more new theories with various intriguing features.

This paper is organized as follows. In Section~\ref{sect:II} we review the basic features of AD theories and their
type IIB constructions. Section~\ref{sect:III} presents the classification of irregular punctures (singularities) in the M5 brane construction, as well as their relation to IIB isolated quasi-homogeneous singularities. Section~\ref{sect:IV} gives a detailed 
study of twisted irregular punctures. In Section \ref{sect:V} we discuss some properties such as spectrum and central charges of the AD theories that we have constructed. 
Finally we conclude in Section~\ref{sect:VI} with a discussion of potential directions.

\section{Generality about Argyres-Douglas Theories}
\label{sect:II}
\subsection{Basic features}
The first Argyres-Douglas type theory was found as the IR theory at a special point on the Coulomb branch of  $\mathcal{N}=2$
pure $SU(3)$ gauge theory \cite{Argyres:1995jj}. At this point, besides the massless photons there are two extra mutually non-local massless dyons, and it was argued 
that the IR theory has to be a strongly-coupled interacting SCFT \cite{Argyres:1995jj,Argyres:1995wt}.  
This theory has no Higgs branch, and has a one dimensional Coulomb branch. 
The Seiberg-Witten curve of this theory can be written as:
\begin{equation}
x^2=z^3+u_1 z+ u_2,
\end{equation}
with Seiberg-Witten differential  $\lambda=x dz$. Since the integral of $\lambda$ along one cycle of  SW curve gives the mass of the BPS particles, its scaling dimension has to be 1, which implies that the scaling dimensions of 
$x,z$ satisfy the condition:
\begin{equation}
[x]+[z]=1.
\end{equation}
By requiring each term in the SW curve to have the same scaling dimension, we find 
\begin{equation}
[x]={3\over 5},~~[z]={2\over 5},~~~[u_1]={4\over 5},~~~[u_2]={6\over 5}.
\end{equation}
$u_1$ has scaling dimension less than one and therefore it is a coupling constant, while $u_2$ is a relevant operator parametrizing the Coulomb branch. 
For a $\mathcal{N}=2$ preserving relevant deformation, the sum of scaling dimensions of coupling constant  $m$ and the relevant operator $u$ has to be equal to 2: $[m]+[u]=2$. 
The distinguished feature of AD theories among the $\cN=2$ SCFTs is that the Coulomb branch operators have fractional scaling dimension and they possess relevant operators.

The original method of locating AD theories on the Coulomb branch of $\cN=2$ gauge theory has been generalized to $SU(2)$ with various flavors in \cite{Argyres:1995xn}, and it is further generalized to 
$\mathcal{N}=2$ theory with general gauge group $G$ and fundamental matter in \cite{Argyres:1995wt,Eguchi:1996vu,Eguchi:1996ds}.  The AD theory from pure $SU(n+1)$ gauge theory can be labeled as  $(A_1, A_n)$ theory, and those 
from $SO(2n)$ gauge theory correspond to  $(A_1, D_n)$, and finally those derived from $E_n$ gauge theory are labeled as $(A_1, E_n)$. These labels denote the shape of the BPS quiver of the 
corresponding SCFTs. 

Over the past decade, there have been many exciting developments in the study of these strongly coupled SCFTs, including the BPS spectrum \cite{Cecotti:2010fi,Xie:2012dw,Xie:2012jd,Xie:2012gd,Maruyoshi:2013fwa,Hori:2014tda,Cordova:2014oxa}, central charges and RG flows \cite{Xie:2013jc}, AGT duality \cite{Gaiotto:2012sf}, bootstrap  \cite{Beem:2014zpa}, superconformal indices \cite{Gadde:2011ik,Buican:2015ina,Cordova:2015nma}, S-dualities \cite{Buican:2014hfa,DelZotto:2015rca} etc.

\subsection{Type IIB construction}\label{IIBcon}
We can engineer a large class of 4$d$ $\cN=2$ SCFTs by considering type IIB string theory on an isolated hypersurface singularity in $\mC^4$ defined by a polynomial:
\ie
W(x_1,x_2,x_3,x_4)=0,
\fe
while decoupling gravity and stringy modes.

Without loss of generality we assume the singular point is the origin and the isolated singularity condition means $dW=0$ if and only if $x_i=0$. The holomorphic 3-form on the three-fold singular geometry is given by
\ie
\Omega={dx^1\wedge dx^2\wedge dx^3\wedge dx^4\over d W }.
\fe
For the resulting 4$d$ theory to be  superconformal, the necessary and sufficient conditions on $W$ are the following \cite{Gukov:1999ya,Shapere:1999xr}:
\begin{enumerate}
	\item There exists a $\mC^*$ action on $W$: a collection of \textit{positive} charges $\{q_i\}$ such that $W(\lambda ^{q_i}x_i)=\lambda W(x_i)$. This is related to the scaling symmetry (or $U(1)_R$ symmetry) of the resulting 4$d$ $\cN=2$ SCFT.\footnote{
		In $\cN=2$ 4$d$ SCFT, the scaling dimension of a Coulomb branch scalar chiral primary is proportional to its $U(1)_R$ charge.} 
	\item The $\mC^*$ charges have to satisfy the condition $\sum q_i>1$.\footnote{
		If in the decoupling limit, we send the string coupling $g_s\rightarrow 0$ while keeping the string scale $\ell_s$ fixed, we would end up with a non-gravitational 4$d$ little string theory (LST) whose holographic dual is described by type II string theory in the background $\mR^{3,1}\times \mR_{\phi} \times (S^1\times LG(W))/\Gamma$ with a suitable GSO projection $\Gamma$ that acts as an orbifold to ensure  4$d$ $\cN=2$ spacetime supersymmetry \cite{Ooguri:1995wj,Giveon:1999zm,Kutasov:2001uf}. Here $\mR_{\phi}$ denotes the $\cN=1$ linear dilaton SCFT with dilaton profile $\varphi=-{Q\over 2}\phi$ and $LG(W)$ the $\cN=2$ Landau-Ginzburg (LG) theory with four chiral superfields $x_i$ and superpotential $W(x_i)$. In particular, the $\mC^*$ action on $x_i$ is identified with the $U(1)_R$ symmetry in the LG model. In the low energy limit ($\ell_s\rightarrow 0$), we recover the 4$d$ SCFT from the LST.
		
		Now the $\cN=1$ linear dilaton theory has central charge $3(1/2+Q^2)$, whereas the $\cN=2$ LG model has central charge $3\sum_i (1-2q_i)$. Consistency of the type II string theory on this background requires the worldsheet theory to have a total central charge of 15 which implies $Q^2=2(\sum_i q_i-1)$ thus $\sum_i q_i-1>0$.
	}
	\label{list}
\end{enumerate}
The full Seiberg-Witten geometry of the 4$d$ SCFT can be derived from the mini-versal (universal deformation with minimal base dimension) deformations of the singularity which take the form \cite{arnold2012singularities}
\begin{equation}
F(x_i,\lambda_a)=W(x_i)+\sum_{a=1}^\mu \lambda_a \phi_a.
\end{equation} 
Here $\phi_a$ is the monomial basis of the local quotient algebra 
\ie\label{locQA}
\cA_W={\mC[x_1, x_2, x_3, x_4]\over \cJ_W}
\fe
where 
\ie
\cJ_W=\la {\partial W\over \partial x_1},{\partial W\over \partial x_2},
{\partial W\over \partial x_3}, {\partial W\over \partial x_4}\ra
\fe
is the Jacobian ideal. 

The complex structure deformations $\lambda_a$ of the singularity correspond to the parameters on the Coulomb branch of the $\cN=2$ 4$d$ SCFT.  The Milnor number $\m\equiv\rank \cA_W$ associated with the singularity captures the rank of the BPS lattice. The BPS particles correspond to D3 branes wrapping special-Lagrangian cycles in the deformed 3-fold. Again one can define a three form $\Omega={dx^1\wedge dx^2\wedge dx^3\wedge dx^4\over d F }$, whose integral
over a special Lagrangian three cycle gives the mass of the BPS particle. Demanding the  integral of $\Omega$ to have mass dimension 1, we deduce the scaling dimension of the deformation parameters as
\ie
&[\lambda_a]=\A(1-q(\phi_a)),
\fe
where $q(\phi_a)$ is the $\mC^*$ charge of monomial $\phi_a$ and
\ie
\A={1\over \sum_{i=1}^4 q_i-1}.
\fe
They capture the Coulomb branch parameters of the 4$d$ $\cN=2$ SCFT. 

Cecotti-Neitzke-Vafa constructed a large class of new AD theories by putting type IIB theory on the following special class of isolated hyper surface singularities \cite{Cecotti:2010fi}:
\begin{equation}
f_G(x_1, x_2)+f_{G'}(x_3,x_4)=0.
\end{equation}
Here $f_G(x,y)$ is a polynomial of the following types 
\ie
f_{A_k}(x,y)& =x^2+y^{k+1} ,
\\
f_{D_{k}}(x,y)& =x^2 y+y^{k-1},
\\
f_{E_6}(x,y)& =x^3+y^4,
\\
f_{E_7}(x,y)& =x^3+xy^3, 
\\
f_{E_8}(x,y)& =x^3+y^5.
\fe
These theories are called $(G, G')$ theories, and the scaling dimensions of various operators have a common denominator \cite{Cecotti:2010fi}
\begin{align}
	& r={1\over 4}{h_G+h_{G'}\over \gcd(h_G/2,h_{G'}/2)},~~&{\rm for}~G,G'=A_1, D_{2n}, E_7, E_8 \nonumber\\
	&r={h_G+h_{G'}\over \gcd(h_G,h_{G'})},~~ &{\rm other ~cases}.
	\label{commondenom}
\end{align}
By looking at the defining data of these theories and reorganizing the monomials parametrizing the deformations, one notices the following obvious equivalences among these theories
\begin{align}
&(G,G^{'})\sim(G^{'},G),\nonumber\\
&(A_1, D_4)\sim (A_2,A_2),~(A_1, E_6)\sim (A_2, A_3),~(A_1, E_8)\sim (A_2, A_4),~\ldots 
\end{align}
The BPS quiver for these theories is given by the direct product of the $G$ and $G'$ type Dynkin diagram. In particular, the dimension of the charge lattice $\Gamma$ is 
\begin{equation}
\dim \Gamma=2n_c+n_f=\rank (G)\times \rank (G^{'}).
\end{equation}
Here $n_c$ is the dimension of Coulomb branch and $n_f$ counts the number of mass parameters. 

In additional to these $(G,G')$ theories, one can engineer a large class of new $\mathcal{N}=2$ SCFTs by classifying the isolated hypersurface singularity with a $\mC^*$ action
satisfying the conditions  $\sum q_i>1$ \cite{danyau}, see \cite{Cecotti:2011gu,DelZotto:2011an} for earlier sporadic examples. We will see below that some of them can be also engineered using M5 brane constructions.\footnote{
	In fact, as we will see, most of the theories in \cite{Cecotti:2011gu,DelZotto:2011an} will turn out to have M5 brane constructions.} 

The connection between the IIB string theory and M5 brane constructions are most transparent in the case of $(A_{k-1},A_{n-1})$ theories \cite{Klemm:1996bj,Cecotti:2010fi}. In that case, IIB string theory on the singular 3-fold $x^k+z^n+y^2+w^2=0$ is T-dual to the IIA NS5 brane wrapped on the singular algebraic curve $x^k+z^n=0$ at $y=w=0$. Lifting to M-theory, we have a M5 brane wrapping the same curve. The deformations $x^i z^j$ of the curve which describe the Coulomb branch of the AD theory are identified with complex structure deformations of the three-fold singularity in the IIB picture. Although such explicit duality transformation is absent in general, we can still argue by comparing the derived Coulomb branch spectrum that a special class of IIB isolated singularities is related to M5 brane configurations.

Recently, there has been an attempt in \cite{Argyres:2007tq,Argyres:2015ffa} to classify $\cN=2$ rank one SCFTs using Kodaira's classification of degeneration of elliptic fibrations. It would be interesting to see if we can find new rank one theories using 3-fold singularities.

\section{Classification of Irregular Singularities}
\label{sect:III}
\subsection{$A$ type irregular singularities}
\label{sectHitchinM5}
One can also engineer four dimensional $\mathcal{N}=2$ SCFTs by putting 6$d$  $A_{N-1}$ $(2,0)$  theory on a Riemann surface $\cC$ with regular (tame) or 
irregular (wild) singularities \cite{Gaiotto:2009we,Xie:2012hs,Chacaltana:2012zy}. 
The SW curve $\Sigma$ of the corresponding field theory can be identified with the spectral curve of the Hitchin system defined 
on $\cC$ \cite{Hitchin:1986vp,Hitchin:1987mz}: 
\ie
\det(x-\Phi)=0\rightarrow x^N+\sum_{i=2}^N \epsilon_i x^{N-i}=0.
\fe
Here $\Phi\in H^0(\cC,End(E)\bigotimes K_\cC)$ is the Higgs field transforming as a holomorphic section of the bundle $End(E)\bigotimes K_\cC$ \footnote{
	The Higgs field $\Phi$ has origin in the 5$d$ $\cN=2$ SYM from $S^1$ reduction of the 6$d$ $(2,0)$ type $J$ theory. Upon twist compactification of the 5$d$ MSYM on $\cC$, a natural principal bundle $E$ arises on $\cC$ with a structure group whose Lie algebra is $J$. Moreover two of the five 5$d$ scalars combine and become the holomorphic section of the bundle $End(E)\times K_\cC$ which is just $\Phi$.
}, and $\epsilon_i$ is the holomorphic section of the line bundle $K_{\cC}^i$. 
Moreover the Seiberg-Witten differential is just $\lambda=xdz$ where $z$ is the holomorphic coordinate on $\cC$.

\begin{figure}[htb]
	\centering
	\includegraphics[scale=0.6]{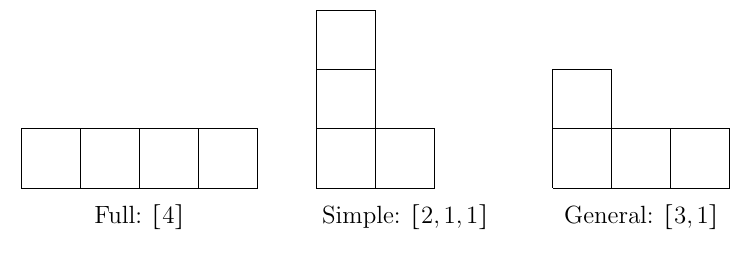}
	\caption{Regular punctures of $A_3$ theories.}
	\label{fig:A3punctures}
\end{figure}

The singularity is characterized by the singular boundary condition of the Higgs field. In particular the regular singularity means that the Higgs field has 
a first order pole:
\begin{equation}
\Phi={T\over z}+\ldots
\end{equation}
where we  have suppressed the regular terms and $T$ is a nilpotent element of the Lie algebra $A_{n-1}$. Using the gauge invariance of the Hitchin system, the regular punctures are classified by 
the nilpotent orbits which can be labeled by a Young tableau $[d_1, d_2, \ldots, d_k]$ \cite{Gaiotto:2009we}. See Figure~\ref{fig:A3punctures} for some examples \footnote{
	We are following the convention of \cite{Chacaltana:2012zy,Chacaltana:2013oka} here. The rows of the Young tableaux correspond to Hitchin partitions which are related to the Nahm partitions via the Spaltenstein map \cite{collingwood1993nilpotent,Chacaltana:2012zy}.}.

One can decorate the
Riemann surface $\cC$ with an arbitrary number of regular singularities. The Coulomb branch chiral primaries of the resulting 4$d$ $\cN=2$ SCFT have integer scaling dimensions and therefore there are no relevant 
chiral primaries in their Coulomb branch spectrum. 

To get an Argyres-Douglas theory, we need to use  irregular singularities. This program has been implemented in \cite{Xie:2012hs} for 6$d$ $A_{n-1}$ theory (see \cite{Gaiotto:2009hg,Cecotti:2011rv,Bonelli:2011aa} for construction in $A_1$ case).  Due to the requirement of superconformal invariance, one
can have only the following two scenarios \cite{Xie:2012hs}: {\bf a}. a single irregular singularity on $\mP^1$; {\bf b}. an irregular singularity and a regular singularity on  $\mP^1$.
The irregular singularities have been classified in \cite{Xie:2012hs}, and they take the following forms:
\ie
& \Phi={T\over z^{r+2}}+\ldots,\quad r={j\over n}>-1,\quad\text{Type \rbI } \nonumber\\
& \Phi={T\over z^{r+2}}+\ldots,\quad r={j\over n-1}>-1,\quad\text{Type \rbII  } \nonumber\\
&\Phi={T_\ell\over z^\ell}+\ldots+{T_1\over z}+\dots,\quad T_\ell\subseteq T_{\ell-1}\subseteq \dots \subseteq T_1,\quad\text{Type \rbIII }
\fe
where we used the usual partial ordering of Young tableau $T_i$ via containment or more generally the partial ordering of associated nilpotent orbits.

\begin{figure}[!htb]
	\centering
	{
		\begin{minipage}{0.45\textwidth}
			\centering
			\includegraphics[scale=.8 ]{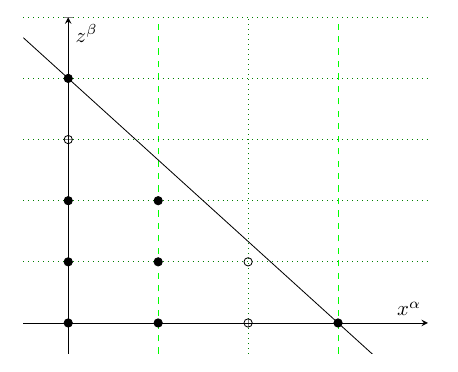}\\
		\end{minipage} 
	}
	\centering
	{
		\begin{minipage}{0.45\textwidth}
			\centering
			\includegraphics[scale=.8 ]{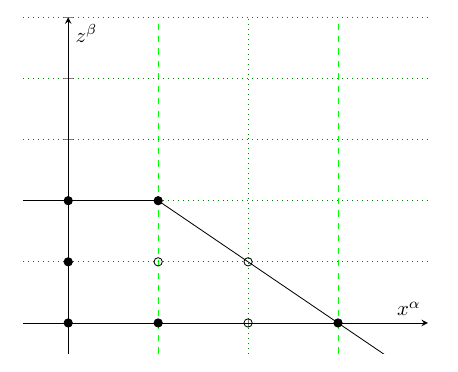}\\
		\end{minipage} 
	}
	\caption{Type I and II irregular singularities in $A_2$ theories with $r={4\over 3}$ and $r={1}$ respectively.}
	\label{fig:A2newton}
\end{figure}

For type I and type II theories, the SW curve can be read from the Newton polygon which captures the leading order behavior of the singularity. Assume the singularity has the following form
\begin{equation}
\Phi\sim \left(\begin{array}{ccc} 
B_1&~&~ \\
~&\ldots&~\\ 
~&~&B_k \\ 
\end{array}\right)
\end{equation}
Here $B_r$ are all diagonal and the order of pole satisfying the condition $r_1<r_2\ldots<r_k$. The size of those blocks sums up to $n$: $d_1+\ldots+d_k=n$.
The Newton polygon is depicted by starting with the point $(n,0)$, and locating a point $(a_i,b_i)$ such that the line connecting the above two points has slope
$-r_i$; next we find another point $(a_{i-1}, b_{i-1})$ such that the subsequent slope is $-r_{i-1}$ etc. See Figure \ref{fig:A2newton} for the Newton polygons of type I and type II singularities. 

Once the Newton polygon is given, one can find out the full 
Seiberg-Witten curve $\Sigma$ by enumerating the integral points contained in the Newton polygon, i.e. we associate to a marked point with  coordinate $(i,j)$ a monomial $x^i z^j$, and the SW curve is simply 
\begin{equation}
\sum_{(i,j)\in S} u_{i,j} x^i z^j=0,
\label{fullSW}
\end{equation}
where the coefficients label the parameters of the Coulomb branch of the AD theory.

One can find the scaling dimensions of these parameters by demanding each term in \eqref{fullSW} to have the same scaling dimension and  that the SW differential $\lambda=x dz$ has scaling dimension 1. Among the parameters of the physical theory, couplings are given by those with $[u_{i,j}]<1$, Coulomb branch operators if $[u_{i,j}]> 1$ and masses if $[u_{i,j}]=1$. For type I and type II theory, some of the deformations are not allowed \footnote{
	This happens when it is a redundant deformation due to coordinate redefinition(s) keeping $\lambda=xdz$ fixed, or because of trace relations for the differentials in the spectral curve.}, see Figure~\ref{fig:A2newton} for the integer points labeled by the  empty dots under 
the Newton polygon.

Now we provide a different justification for why there are only type I and type II irregular singularities. At the origin of the Coulomb branch moduli space, the SW curves for type I and type II theories are 
\begin{align}
	&x^n+z^k=0~~~\text{Type \rbI }, \nonumber \\
	& x^n+ x z^{k}=0~~~\text{Type \rbII  } .
\end{align}
The spectral curve of the $A_{n-1}$ type Hitchin system may be written as the three-fold form 
\begin{equation}
x_1^2+x_2^2+x_3^n+\sum_{i=2}^n\epsilon_i(z) x_i^{n-i}=0.
\label{typeA3foldform}
\end{equation}
Here $\epsilon_i(z)\in K^i$ and is a polynomial in $z$.  We would like to have an isolated singularity at the origin, and the only two possibilities are the following (see Appendix~\ref{apx:isocdv}):
\begin{align}
	&x_1^2+x_2^2+x_3^n+z^k=0,~~~\text{Type \rbI }, \nonumber \\
	& x_1^2+x_2^2+x_3^n+x z^k=0.~~~\text{Type \rbII  } .
	\label{typeAiibsing}
\end{align}
Forgetting the first two quadratic terms which are rigid, we see that the classification of the irregular singularities in the $A$ type Hitchin system boils down to, in the IIB perspective, two types of isolated  three-fold 
hypersurface singularities in \eqref{typeAiibsing} among the ones with the form of \eqref{typeA3foldform}.

\subsection{General case}
\label{classifyirregsing}
Now we would like to generalize the classification of irregular singularities to other types of 6$d$ $(2,0)$ theory labeled by a Lie algebra $J=D,E$. 
We still use $z$ to denote the coordinate on $\mP^1$ and declare that the Higgs field has the following form
near $z=0$,
\ie
\Phi=B+\dots.
\label{generalsingularity}
\fe
Here $B$ is the singular term which can be put in the following block-diagonal form:
\begin{equation}
\Phi\sim \left(\begin{array}{ccc} 
B_1&~&~ \\
~&\ldots&~\\ 
~&~&B_k \\ 
\end{array}\right)+\ldots
\label{singularityblockdiagonal}
\end{equation}
with the order of pole for various blocks ordered as $r_1<r_2<\ldots<r_k$. 

For $D$ type theory we use the fundamental representation, while for $E$ type theory we use  the adjoint representation for the polar matrices appearing in the above description. 
The physics should not depend on the representation we are using \footnote{
	One need to be careful about the constraints among the differentials appearing in the spectral curve when working with a general representation.}.

We determine what kind of combination of $r_i$ and forms of $B_i$ are needed to define a SCFT.  
This may be done using a similar method which has been used in \cite{Xie:2012hs} for the type $A$ case. 

\begin{table}
	\begin{center}
		\begin{tabular}{ |c|c| c|c|c|c| }
			\hline
			$J$ & 
			$\dim J$ & $h$  & $\{d_i\}_{i=1,\dots,\,\rank(J)}$\\ \hline
			$A_{n-1} $  & $n^2-1$ & $n$ & $2,3,\dots,n$ \\     \hline
			$D_n$   & $n(2n-1)$ & $2n-2$ & $2,4,\dots,2n-2;n$ \\     \hline
			$E_6$  & $78$ & $12$ & $2,5,6,8,9,12$\\     \hline
			$E_7$  & $133$ & $18$ &$2,6,8,10,12,14,18$ \\     \hline
			$E_8$   & $248$ & $30$ & $2,8,12,14,18,20,24,30$ \\     \hline
		\end{tabular}
	\end{center}
	\caption{Relevant Lie algebra data: $h$ denotes the Coxeter number and $\{d_i\}$ are degrees of the fundamental invariants .}
	\label{liedata}
\end{table}

However, given the correspondence between $A$ type irregular singularities in the Hitchin system on $\cC$ from the M5 brane perspective and isolated singularities of the form \eqref{typeAiibsing} in IIB geometry, we find it more convenient to generalize the classification of $A$ type irregular singularities in the language of IIB isolated singularities. 


To begin with, let us review some properties of the $D$ type and $E$ type Hitchin system. For $D$ type theories, the SW curve looks like 
\begin{equation}
x^{2n}+ \sum_{i=1}^{n-1}\epsilon_{2i}(z) x^{2n-2i}+ (\tilde\epsilon_n(z))^2=0.
\end{equation}
Here $\epsilon_{2i} \in K^{2i}$ for $i=1,\dots,n-1$ and $\tilde\epsilon_n(z)\in K^n$. The novelty here compared to the type $A$ discussion is that the term constant in $x$ is constrained to be a perfect square. The coefficients in these differentials parametrize the Coulomb branch of the 4$d$ SCFT.  
The spectral curve for the $E$ type Hitchin system is much more complicated due to constraints among the differentials. Here the important fact for us is that the independent invariant polynomials 
parametrizing the Coulomb branch are 
\begin{align}
	&E_6:~~~ \epsilon_2(z),~ \epsilon_5(z),~ \epsilon_6(z),~ \epsilon_8(z),~ \epsilon_9(z),~ \epsilon_{12}(z), \nonumber\\
	&E_7:~~~ \epsilon_2(z),~\epsilon_6(z),~\epsilon_8(z),~\epsilon_{10}(z),~\epsilon_{12}(z),~  \epsilon_{14}(z),~\epsilon_{18}(z),\nonumber\\
	&E_8:~~~ \epsilon_2(z),~\epsilon_{8}(z),~\epsilon_{12}(z),~\epsilon_{14}(z),~\epsilon_{18}(z),~\epsilon_{20}(z), ~\epsilon_{24}(z),\nonumber\\
		&~~~~~~~~~\epsilon_{30}(z).
\end{align}
The above differentials are holomorphic sections of various line bundles $\epsilon_i(z)\in K^i$ over $\cC$. 
To utilize IIB description, one can put the SW curve in the three-fold form \cite{Klemm:1996bj}:
 
\begin{align}
	&A_{n-1}:~x_1^2+x_2^2+ x_3^n+\epsilon_2(z)x_3^{n-2}+\ldots+\epsilon_{n-1}(z)x_3+\epsilon_{n}(z) =0,  \nonumber\\
	&D_n:~x_1^2+x_2^{n-1}+x_2 x_3^2+\epsilon_2(z)x_2^{n-2}+\ldots+\epsilon_{2n-4}(z)x_2+\epsilon_{2n-2}(z)+\tilde{\epsilon}_n(z)x_3=0,  \nonumber\\
	&E_6:~x_1^2+x_2^3+x_3^4+\epsilon_2(z)x_2x_3^2+\epsilon_5(z)x_2x_3+\epsilon_6(z)x_3^2+\epsilon_8(z)x_2+\epsilon_9(z)x_3+\epsilon_{12}(z)=0, \nonumber\\
	&E_7:~x_1^2+x_2^3+x_2x_3^3+\epsilon_2(z)x_2^2x_3+\epsilon_{6}(z) x_2^2+\epsilon_{8}(z)x_2x_3+\epsilon_{10}(z)x_3^2   \nonumber\\
	&~~~~~~+\epsilon_{12}(z)x_2+\epsilon_{14}(z)x_3+\epsilon_{18}(z)=0, \nonumber\\
	& E_8:~x_1^2+x_2^3+x_3^5+\epsilon_2(z)x_2x_3^3+\epsilon_8(z)x_2x_3^2+\epsilon_{12}(z)x_3^3+  \nonumber\\
	&~~~~~~~\epsilon_{14}(z)x_2x_3+\epsilon_{18}(z)x_3^2+\epsilon_{20}(z)x_2+\epsilon_{24}(z)x_3+\epsilon_{30}(z)=0. 
	\label{3foldforms}
\end{align}
 
Now $\epsilon_i(z)$ are polynomials in $z$, and we would like to find out the choice of $\epsilon_i$ to turn on such that there is an isolated singularity at the origin.

Three-fold hypersurface singularities of the form \eqref{3foldforms} are called compound Du Val (cDV) singularities in singularity theory \footnote{
	The compound Du Val (cDV) singularities are a special class of 3-fold  singularities defined by 
	\ie
	W_J(x_1,x_2,x_3,z)=f_J(x_1,x_2,x_3)+zg(x_1,x_2,x_3,z)=0
	\label{cDV}
	\fe
	where $f_J$ is the usual Du Val (DV) singularity of $J=A,D,E$ type and $g(x_1,x_2,x_3,z)$ is an arbitrary polynomial. The case in which $g(x_1,x_2,x_3,z)=z$, the cDV singularities reduce to 3-fold Du Val singularities.

	Demanding the $\mC^*$ charge of $z$ to satisfy $q(z)>0$, using coordinate redefinitions, we can put any cDV singularity into the form of \eqref{3foldforms}.
}. 
\begin{table}[!htb]
	\begin{center}
		\begin{tabular}{ |c|c|c|c| }
			\hline
			~& \shortstack{Singularity\\ {}} & \shortstack {\\Leading order \\differential} &\shortstack{ \\Label\\{}}   \\ \hline
			$A_{n-1}$ &$x_1^2+x_2^2+x_3^n+z^k=0$&$\epsilon_n=z^k$& $A_{n-1}^{(n)}[k]$\\ \hline
			$~$& $x_1^2+x_2^2+x_3^n+ z^kx_3=0$ &$\epsilon_{n-1}=z^k$& $A_{n-1}^{(n-1)}[k]$\\ \hline
			
			$D_n$   & $x_1^2+x_2^{n-1}+x_2x_3^2+z^k=0$ & $\epsilon_{2n-2}=z^k$ & $D_{n}^{(2n-2)}[k]$ \\     \hline
			$~$   &$x_1^2+x_2^{n-1}+x_2x_3^2+z^k x_3=0$&$\tilde{\epsilon}_{n}=z^k$ &  $D_{n}^{(n)}[k]$ \\     \hline
			
			$E_6$  & $x_1^2+x_2^3+x_3^4+z^k=0$&$\epsilon_{12}=z^k$ & $E_6^{(12)}[k]$   \\     \hline
			$~$  & $x_1^2+x_2^3+x_3^4+z^k x_3=0$ &$\epsilon_{9}=z^k$ &$E_6^{(9)}[k]$   \\     \hline
			$~$  & $x_1^2+x_2^3+x_3^4+z^ k x_2=0$  &$\epsilon_{8}=z^k$ &$E_6^{(8)}[k]$   \\     \hline
			
			$E_7$  & $x_1^2+x_2^3+x_2x_3^3+z^k=0$& $\epsilon_{18}=z^k$& $E_7^{(18)}[k]$    \\     \hline
			$~$  & $x_1^2+x_2^3+x_2x_3^3+z^kx_3=0$ & $\epsilon_{14}=z^k$&$E_7^{(14)}[k]$    \\     \hline

			$E_8$   & $x_1^2+x_2^3+x_3^5+z^k=0$&$\epsilon_{30}=z^k$ &$E_8^{(30)}[k]$   \\     \hline
			$~$   & $x_1^2+x_2^3+x_3^5+z^k x_3=0$ &$\epsilon_{24}=z^k$&$E_8^{(24)}[k]$   \\     \hline
			$~$   & $x_1^2+x_2^3+x_3^5+z^k x_2=0$ & $\epsilon_{20}=z^k$&$E_8^{(20)}[k]$  \\     \hline
			
		\end{tabular}
	\end{center}
	\caption{Isolated quasi-homogeneous cDV singularities.}
	\label{table:isoscDV}
\end{table}

It is straightforward to prove (see Appendix~\ref{apx:isocdv} for details) that
\begin{tcolorbox}
	The isolated quasi-homogeneous 3-fold singularities of cDV type are precisely the ones listed in Table~\ref{table:isoscDV}.
\end{tcolorbox}

We find the form of the irregular singularity for the Higgs field $\Phi$ on $\cC$ such that the leading order differential (which defines the singularity) is given by the terms listed in Table~\ref{table:isoscDV}. In other words, we want to identify the Hitchin system that describes the same Coulomb branch spectrum of some $\cN=2$ SCFT as does the IIB singular geometry.

\begin{figure*}[!htb]
	\centering
\includegraphics[scale=.9]{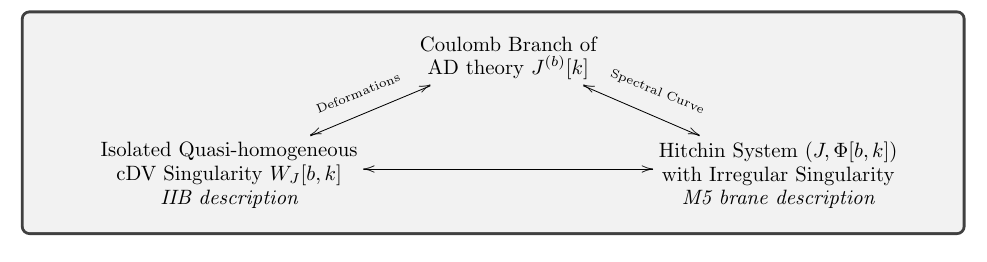}
\caption{Connection between AD theory, cDV singularity, and Hitchin system.}
	\label{fig:ADsh}
\end{figure*}

It is straightforward to check by explicitly comparing the Coulomb branch parameters from the spectral curve of the Hitchin system with those from the complex structure deformations of the 3-fold singularity (see subsection~\ref{DtypeADuntwisted} for an illustration in $D$-type theories) that the Higgs field has the following singular form 
\begin{equation}
\Phi_z={T\over z^{2+k/b}}+\dots,
\end{equation}
where $T$ is a regular semi-simple element of the Lie algebra $J$ and the possible values of  $b$ are listed in Table~\ref{table:slopes1} which are in one to one correspondence with the degrees of the leading order differentials in Table~\ref{table:isoscDV} \footnote{
	Up to conjugation, $J$ has infinitely many regular semi-simple elements. Here the choice of $T$ is constrained by $b$ to ensure gauge invariance across the branch cuts on the $z$ plane.}. Those are the irregular singularities that we  focus on in this paper and we denote the resulting $\cN=2$ SCFT by $J^{(b)}[k]$. We summarize the connections between the AD theory and its various descriptions in Figure~\ref{fig:ADsh}.

%

When the allowed denominator $b$ is taken to be the dual Coxeter number $h$ of the corresponding Lie algebra $J$, the 4$d$ $\cN=2$ SCFT $J^{(h)}[k]$ engineered using this singularity corresponds to the $(J, A_{k-1})$ theory of \cite{Cecotti:2010fi} reviewed in the previous section. Moreover, most of the theories constructed in \cite{Cecotti:2011gu,DelZotto:2011an} using  Arnold's unimodal and bimodal singularities \cite{arnold2012singularities} are also included in our construction \footnote{
	The only exceptions are $Z_{12},S_{12}$ in the unimodal case, and $Z_{18},Q_{17},S_{16}$ for the bimodal case.}.


Interestingly Table~\ref{table:slopes1} also shows up in the discussion of {\it non-degenerate} \footnote{
	This notion of non-degeneracy is consistent with the requirement of isolated singularities that gives rise to SCFTs, which restricts to the case of a single block in \eqref{singularityblockdiagonal}.
}
Hitchin systems with irregular singularities for which the Higgs field has a leading polar matrix of the regular semi-simple type (go to a covering space of the $z$ plane if necessary) \cite{Witten:2007td,2009arXiv0901.2163F}. We start with a irregular singularity of the type
\ie
\Phi={T\over z^n}+\cdots,\quad n\in \mZ,~n>1.
\fe
When $T$ is already regular semi-simple on the $z$ plane, the story is straightforward and we do not have further constraints on $T$. On the other hand if $T$ is not semisimple, for example nilpotent, it is well known that after lifting to a $b$-fold cover of the local patch around the singularity on the $z$ plane by $z=u^b$, \eqref{generalsingularity} is gauge equivalent to a local model with leading polar matrix $T'$ semisimple and leading singularity of order $a+1$ \cite{2009arXiv0901.2163F} \footnote{
	As an example \cite{2009arXiv0901.2163F}, let us fix a Borel subalgebra of $J$ and denote the Lie algebra element corresponding to the root $\A$ by $X_\A$. We then define nilpotent elements $N=\sum_{\A\in \Delta} X_\A$ and $E=X_{-\theta}$ where $\Delta$ is the set of simple roots and $\theta$ is the longest (positive) root. The following local model for the Higgs field,
	\ie
	\Phi\sim {E\over z^2}+{N\over z}+\dots
	\fe
	is gauge equivalent to (after projecting down to the $z$ plane)
	\ie
	\Phi\sim h{E+N\over z^{1+1/h}}+\dots
	\fe
	where $E+N$ is regular semisimple and $h$ is the Coxeter number.
}.
The local model written in terms of the original coordinate $z$ is \footnote{
	The presence of ``branch cuts'' on the $z$-plane indicates nothing but a wrong choice of complex structure. Recall that the Hitchin equations are also invariant under non-holomorphic gauge transformations which we can make use of to remove the ``branch cuts'' \cite{Witten:2007td}.
}
\ie
\Phi={T'\over z^{1+{a\over b}}}+\cdots.
\label{localmodelslope}
\fe
The ratio $s=a/b$ (we take $b$ to be the minimal possible integer possible) is called the \textit{slope} or Katz invariant associated with the local model of the Higgs field \cite{2009arXiv0901.2163F,2014arXiv1404.0598C}. It can be shown using the relation between Higgs bundles and opers that the denominator $b$ of $s$ must always be a divisor of $d_i$ for some $1\leq i\leq \rank (J)$, which are degrees of the fundamental invariants of $J$ (see Table \ref{liedata})  \cite{2014arXiv1404.0598C}. Further imposing that $T'$ is regular semisimple restricts the denominator $b$ of the slope to take the values summarized in Table~\ref{table:slopes1}.\footnote{
	This follows from a classification result in the work of Springer \cite{springer1974regular} and later Kac et al \cite{elashvili2013cyclic}. More explicitly,
	the non-single-valuedness of the Higgs field around the origin on the $z$-plane demands a nontrivial gauge transformation $g$ across the branch cuts,
	\ie
	\Phi(ze^{2\pi i})=g\Phi(z) g^{-1}=\omega^a \Phi(z)
	\fe
	where $\omega$ is the $b$-th root of unity.  
	In general $g$ can be identified with a element $w$ of the Weyl group $W(J)$ and $T'$ an eigenvector of $g$ (or $w$) which lies on a plane in the Cartan subalgebra $\mft_{\mC}$ fixed by $w$. Requiring $T'$ to be regular semisimple implies $w$ is a regular element in the sense of \cite{springer1974regular} which was classified and their orders correspond to the allowed $b$s (we omit the $b$'s which are divisors of other ones). Related to this, $T'$ is also what is called a regular semisimple cyclic element in \cite{elashvili2013cyclic}, which has also been classified and the list again coincide with that of Table~\ref{table:slopes1}.
}
For example, for $A_n$ type theories, the possible denominators for the slopes are simply $n+1$ and $n$ which correspond to the Type I and II theories described previously in \cite{Xie:2012hs}.

\subsubsection{Maximal irregular singularity}
\label{sect:maximalut}
A special class of AD theories can be constructed from irregular singularity of the maximal type: namely the Higgs field behaves as
\begin{equation}
\Phi={T_\ell \over z^\ell}+{T_{\ell-1}\over z^{\ell-1}}+\dots+{T_1\over z}+\cdots.
\end{equation}
Here $T_i$ are in regular semi-simple orbits of $J$. This corresponds to the case where $k$ is a positive integer multiple of $b$ in \eqref{localmodelslope}. The dimension of the Coulomb branch is \cite{Witten:2007td}
\ie
\dim {\rm Coulomb}={\ell(\dim J -\rank J)\over 2}-\dim J.
\fe
The number of mass parameters of this theory is $n_f=\rank J$, and therefore
the dimension of the BPS charge lattice is 
\ie
\dim \Gamma=&2\dim {\rm Coulomb}+n_f
\\
=&\ell(\dim(J )-\rank J)-2\dim(J)+\rank J
\\
=&\rank J [(\ell-2)h(J)-1]
\fe
This corresponds to the $(J,A_{(\ell-2)h(J)-1})$ theory of \cite{Cecotti:2010fi}.

We expect this class of theories to have a rich set of features among all AD theories (some of which we exhibit in section~\ref{sect:featMAX}). They typically have Higgs branches, large flavor symmetries and are likely to have 3d mirror quiver gauge theories \cite{Benini:2010uu,Xie:2012hs}. There exist exactly marginal operators in the Coulomb branch spectrum and the theory can undergo nontrivial S-duality transformations. 

\subsubsection{Degeneration of irregular singularities} 
\label{sect:degenut}
Let us now consider a degeneration of the irregular singularity considered in the previous subsection.
\begin{equation}
\Phi={T_\ell \over z^\ell}+{T_{\ell-1}\over z^{\ell-1}}+\ldots+{T_1\over z}+\ldots 
\end{equation}
which is specified by a sequence of semisimple elements: $\rho=\{T_1,T_2,\dots,T_\ell\}$ of $J$.
Previously we took all of these matrices from the regular semi-simple orbits. In general, we could take them to be from other semisimple orbits, and the only constraints are \cite{Xie:2012hs,Witten:2007td}
\begin{equation}
T_1 \subseteq T_2\subseteq\ldots \subseteq T_\ell. 
\end{equation}
For this type of puncture, the local contribution of the singularity $\rho$ to the dimension of the Coulomb branch is 
\begin{equation}
\dim_{\rho} {\rm Coulomb} ={1\over 2}\sum_{i=1}^\ell \text{dim}(T_i),
\end{equation}
where $\text{dim}(T_i)$ is the (complex) dimension of the corresponding orbit. Moreover, the number of the mass parameters in the resulting AD theory from this singularity is equal to the number of distinguished eigenvalues of $T_1$. 

Some $\cN=2$ SCFTs constructed using this type of singularity have exactly marginal operators and nontrivial S-duality. One example of such has been explored in \cite{Buican:2014hfa}.

\subsection{Some Explicit Examples of AD Theories}

\subsubsection{$D$ type AD theories and Newton polygon}
\label{DtypeADuntwisted}
\begin{figure}[htb]
	\centering
	{
		\begin{minipage}{0.45\textwidth}
			\centering
			\includegraphics[scale=.8 ]{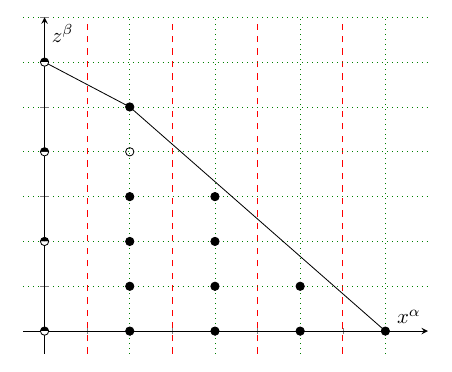}\\
		\end{minipage} 
	}
	\centering
	{
		\begin{minipage}{0.45\textwidth}
			\centering
			\includegraphics[scale=.8 ]{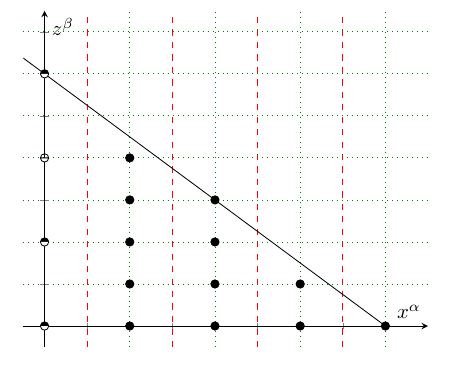}\\
		\end{minipage} 
	}
	\caption{Newton polygons for  $D_4^{(6)}[5]$ and $D_4^{(4)}[6]$ theories.}
	\label{fig:type1&2D4theories}
\end{figure}

The SW curve for the $D$ type theory can be easily read off from the spectral curve of the Hitchin system,  
\begin{equation}
x^{2n}+ \sum_{i=1}^{n-1}\epsilon_{2i}(z) x^{2n-2i}+ (\tilde\epsilon_n(z))^2=0.
\end{equation}
There are two types of  AD theories $D_n^{(n)}[k]$ and $D_n^{(2n-2)}[k]$ which correspond to the following singular SW curves 
\begin{align}
	& x^{2n}+ z^{2k}=0 ,  \nonumber\\
	& x^{2n}+z^k x^2=0,
\end{align}
with the SW differential $\lambda=xdz$.
The Higgs field takes the following singular forms accordingly
\ie
&\Phi={T\over z^{2+{k\over n}}}+ \cdots , 
\\
&\Phi={T\over z^{2+{k\over 2n-2}}}+ \cdots .
\fe
One can read off the scaling dimension of $x$ and $z$ by requiring that the 
SW differential has scaling dimension 1: $[x]+[z]=1$. The full SW curve can be easily found from the Newton polygon, see Figure~\ref{fig:type1&2D4theories}. 

Let us for illustration consider the example $D^{(6)}_4[5]$. We can write down the full SW curve using the Newton polygon (Fig.~\ref{fig:type1&2D4theories}) as follows. We list monomials $x^\A z^\B$ that correspond to filled dots in the Newton polygon. The half-filled dots on the $x^0$ axis indicate that we should only regard the square root of the corresponding monomial $z^{\B}$ as parametrizing independent deformations. This is due to the Pfaffian constraint. Hence we have
\ie
&x^8+x^6(u_{1,1} z+u_{1,0})+x^4(u_{2,3}z^3+u_{2,3}z^2+u_{2,1}z+u_{2,0})
\\
&
+x^2(z^5+u_{3,3}z^3+u_{3,2}z^2+u_{3,1}z+u_{3,0})
\\
& +(\tilde u_3 z^3+\tilde u_2 z^2+\tilde u_1 z+\tilde u_0)^2=0
\fe
From the scaling dimensions of $x$ and $z$,
\ie
&[x]={5\over 11},\quad [z]={6\over 11},
\fe
we can read off the dimensions of the Coulomb branch parameters. In particular there are no mass parameters and the chiral primaries have dimensions
\ie
\Delta_{\rm Coulomb}=\{{12\over 11},{14\over 11},{14\over 11},{18\over 11},{20\over 11},{20\over 11},{24\over 11},{30\over 11}\}
\label{D6k=4CB}
\fe
among which the relevant ones are paired with coupling constants as expected.

We can equivalently use the IIB description with the isolated hypersurface singularity
\ie
W(x,y,z,w)=w^2+x^3+xy^2+z^5=0
\fe
whose local quotient algebra is
\ie
\cA_W=&\{1,x,y,z,xz,y^2,yz,z^2,xz^2,y^2z,yz^2,z^3,
\\
&
y^2z^2,xz^3,yz^3,y^2z^3\}.
\fe
From the $\mC^*$ charges of the coordinates
\ie
&q_x={1\over 3},~q_y={1\over 3},~q_z={1\over 5},~q_w={1\over 2} 
\fe
and
\ie
\A={1\over \sum_{q_i}-1}={30\over 11}
\fe
we recover the same Coulomb branch spectrum as in \eqref{D6k=4CB}.

\subsubsection{An irregular singularity and an regular singularity}
To build AD theories with generically non-Abelian flavor symmetries, we can consider an irregular singularity and a regular singularity on $\mP^1$. The regular singularity is labeled by 
a nilpotent orbit of $J$ (the Nahm description (or Higgs branch description) is better). See \cite{Chacaltana:2012zy} for various types of punctures. There are a variety of new theories by choosing different 
regular and irregular punctures.

\begin{figure}[htb]
	\centering
	{
		\begin{minipage}{0.45\textwidth}
			\centering
			\includegraphics[scale=.8 ]{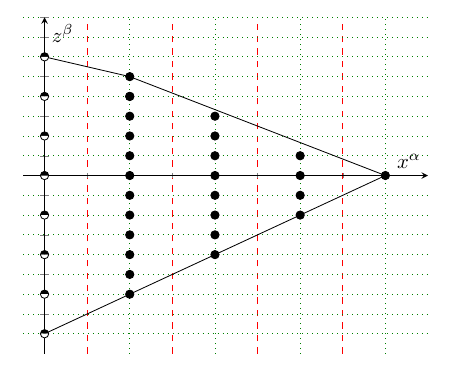}\\
		\end{minipage} 
	}
	\centering
	{
		\begin{minipage}{0.45\textwidth}
			\centering
			\includegraphics[scale=.8 ]{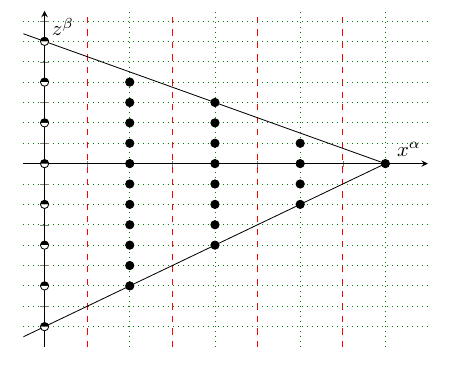}\\
		\end{minipage} 
	}
	\caption{Newton Polygons for $(D_4^{(6)}[5],F)$ and $(D_4^{(4)}[6],F)$ theories.}
	\label{fig:type4$d$4theories}
\end{figure}

If we take the irregular punctures with pole order denominator $b$ given by the Coxeter number $h$ of the Lie algebra $J$,
\ie
\Phi\sim {T\over z^{2+k/h}}+\cdots
\fe  and choose the full regular puncture, we can  construct AD theories with non-Abelian $J$ flavor symmetry which we denote by $(J^{(h)}[k],F)$ (see Figure~\ref{fig:type4$d$4theories} for their Newton polygons). These 
theories are called $D_p(J)$ theories in \cite{Cecotti:2012jx,Cecotti:2013lda} with $p=k+h$.

\section{Twisted Irregular Singularity}\label{sect:IV}

If the underlying Lie algebra $J$ has a nontrivial outer-automorphism group ${\rm Out}(J)$ (see Table \ref{table:outerautomorphisms}, it induces an automorphism on the Hitchin moduli space. Therefore, we may consider the projection onto ${\rm Out}(J)$ invariant configurations of the Higgs field $\Phi$. This can be done locally at the singularities via introducing monodromy twist by an element $o\in  {\rm Out}(J)$,
\ie
\Phi(e^{2\pi i }z)=g[o(\Phi(z))]g^{-1}
\label{condontwisted}
\fe
for some $g\in   J /\mfg^\vee$ (here $\mfg^\vee$ is the invariant subalgebra of $J$), 
which we refer to as \textit{twisted singularities}. Globally, the twisted singularities must come in pairs connected by twist lines (or cuts).

\begin{table}[htb]
	\begin{center}
		\begin{tabular}{ |c|c| c|c|c|c| }
			\hline
			$ J $ ~&$A_{2N}$ &$A_{2N-1}$ & $D_{N}$  &$E_6$&$D_4$ \\ \hline
			Automorphism ${\rm Out}( J )$ &$\mZ_2$ &$\mZ_2$& $\mZ_2$  & $\mZ_2$&$\mZ_3$\\     \hline
			Invariant subalgebra  $\mfg^\vee$ &$B_N$&$C_N$& $B_{N-1}$  & $F_4$&$G_2$\\     \hline
			Langlands dual    $\mfg$ &$C_N$  &$B_N$& $C_{N-1}$  & $F_4$&$G_2$\\     \hline
		\end{tabular}
	\end{center}
	\caption{Outer-automorphisms of simple Lie groups \cite{fulton1991representation}.}
	\label{table:outerautomorphisms}
\end{table}

\subsection{Review of regular twisted singularities}
Twisted singularities of the regular type from which one builds usual Class S theories have been explored extensively in  \cite{Chacaltana:2012zy}. From Table \ref{table:outerautomorphisms}, we see a nontrivial $o$ can have order 2 (for $J=A_{2n-1},D_n,E_6$), or order 3 (for $J=D_4$) \footnote{
	As was pointed out in \cite{Tachikawa:2011ch,Chacaltana:2012zy}, the case $J=A_{2n}$ is subtle due to a discrete theta angle in the 5$d$ maximal SYM from the compactification of the 6$d$ $(2,0)$ type $A_{2n}$ theory on $S^1$ with $\mZ_2$ twist. However it is unclear to us whether this subtlety affects the local irregular singularities discussed here.
}.
The Lie algebra $J$ acquires a grading with respect to the eigenvalues of $o$,
\ie
J=
\begin{cases}
	J_1+J_{-1} & {\rm for~} |o|=2
	\\
	J_1+J_{\omega}+J_{\omega^2} & {\rm for~} |o|=3
\end{cases}
\fe
where $\omega^3=1$ and we use the subscript on $J$ to denote the eigen-subspaces.

The local model for the Higgs field near the singularity takes the following form when $o$ has order $2$,
\ie
\Phi={T_1\over z}+{U_1\over z^{1/2}}+T_0+\dots
\fe
where $T_1,T_0\in J_1= \mfg^\vee$ and $U_1\in J_{-1}$. When the twist element $o$ has order 3 with eigenvalue $\omega$ (third root of unity), we have instead
\ie
\Phi={T_1\over z}+{U_1\over z^{2/3}}+{W_1\over z^{1/3}}+T_0+\dots
\fe
where $T_1,T_0\in J_1=\mfg^\vee$, $U_1\in J_\omega$ and $W_1\in J_{\omega^2}$. 

The Coulomb branch of the resulting 4$d$ SCFT receives local contributions from not only the Spaltenstein-dual nilpotent orbit $d(\cO_{\bf p})$ in $\mfg^\vee$ associated with the leading polar matrix $T_1$ \footnote{
	Here {\bf p} labels the Nahm pole \cite{Chacaltana:2012zy}.}, but also  components of the subleading polar matrices $U_1,W_1$. Altogether the local contribution to the Coulomb branch has dimension \cite{Chacaltana:2012zy}
\ie 
\dim_{{\bf p}} {\rm Coulomb}  ={1\over 2}\dim_{\mC} d(\cO_{\bf p}) +{1\over 2}\dim J/\mfg ^\vee.
\label{twistedregularlocalCB}
\fe
To obtain the total Coulomb branch dimension, we add up the local contributions from the singularities and the global contribution \footnote{
	By Riemann-Roch theorem, the moduli associated with a $j$-th holomorphic differential on a genus $g$ Riemann surface $\cC$ has dimension $
	\dim H^0(K_\cC^j)=\deg(K_\cC^j)-(g-1)=(2j-1)(g-1)
	$. Including contributions from each fundamental invariant differentials and using the Lie-algebraic formula $\dim( J )=2\sum_{i=1}^r d_i-r$, we end up with the total global contribution $(g-1)\dim  J $.
},
\ie
\dim {\rm Coulomb} =\sum_i \dim_{{\bf p}_i} {\rm Coulomb}+(g-1)\dim  J ,
\label{twistedregularglobalCB}
\fe
where $g$ is the genus of the Riemann surface $\cC$.

On the other hand, the local contribution to the Higgs branch has quaternionic dimension  \cite{Chacaltana:2012zy}
\ie
\dim_{{\bf p}}{\rm Higgs} ={1\over 2}\left(\dim \mfg-\rank \mfg-\dim_{\mC} \cO_{\bf p}\right).
\fe
The total quaternionic dimension of the Higgs branch is given by
\ie
\dim {\rm Higgs}=\sum_i \dim_{{\bf p}_i}{\rm Higgs} +\rank \mfg^\vee.
\fe

\subsection{Maximal twisted irregular singularities}
\label{sect:maxtwist}
We now extend the twisted singularities to the irregular type which can be achieved by decorating the irregular singularities considered  previously with appropriate local monodromy twist $o\in {\rm Out}( J )$. As explained before, demanding conformal invariance and $o$-invariance, we specialize to the case of one irregular twisted singularity and one regular twisted singularity on $\mP^1$. Unlike the untwisted case, we do not have a classification for these twisted irregular singularities at the moment. Nonetheless we see a number of subclasses can already be constructed easily and have interesting features. We will leave the general classification of twisted singularities that give rise to AD theories to a future publication. Since IIB description for these twisted singularities is not known, it would also be interesting to figure out the corresponding IIB 3-fold singular geometry.

First let us consider the case where the irregular singularity is of the maximal type discussed in subsection~\ref{sect:maximalut} with a $\mZ_2$ twist. The
local structure of irregular singularity is,
\ie
\Phi={T_\ell\over z^\ell}+{U_\ell\over z^{\ell-1/2}}+{T_{\ell-1}\over z^{\ell-1}}+{U_{\ell-1}\over z^{\ell-3/2}}+\ldots+{T_1\over z}+\dots
\fe
where $T_i$ are regular semisimple elements of $ J_{1}=\mfg^\vee $ which is even under the $\mZ_2$ twist and $U_i\in J_{-1}$ is odd. We denote the data defining the twisted irregular singularity collectively by $\tilde{\rho}=\{T_i,U_j|1\leq i\leq \ell,2\leq j\leq \ell\}$.

The local contribution to the Coulomb branch dimension can be obtained by studying the pole structure of the differentials $\epsilon_{d_i}$. Expanded in $z$, if the leading singular term in $\epsilon_{d_i}$ that is not completely determined by the singular part of $\Phi$ (i.e. $T_m$ for $1\leq m\leq \ell$ and $U_m$ for $2\leq m\leq \ell$) has pole order $p_{d_i}$, the irregular singularity contributes $\sum_{i=1}^{\rank J} p_{d_i}$ to the Coulomb branch dimension. Taking into account the local contribution from the regular twisted singularity \eqref{twistedregularlocalCB}, and the global contribution (with now $g=0$ this is $-\dim  J $), we obtain the total Coulomb branch dimension.

The number of distinctive eigenvalues of $T_1$ corresponds to the number of mass parameters contributed by the twisted irregular singularity, which is the rank of $\mfg^\vee$ for the maximal case we consider here. We expect the local contribution to the Higgs branch also has quaternionic dimension $\rank \mfg^\vee$.

For example consider a $\mZ_2$ twist of the $ADE$ type Hitchin system with regular semisimple polar matrices $T_m$. The set of fundamental invariants splits under the action $o$,
\ie
&A_{2n-1}~ o:\epsilon_k\rightarrow (-1)^k\epsilon_k\quad{\rm for}~2\leq k\leq 2n
\\
&D_{n\neq 4}~ o:\epsilon_{2i}\rightarrow \epsilon_{2i}\quad{\rm for}~1\leq i\leq n-1,~{\rm and}~\tilde{\epsilon}_n\rightarrow -\tilde{\epsilon}_n
\\
&E_{6}~o:\epsilon_{i}\rightarrow \epsilon_{k}\quad{\rm for}~k=2,6,8,12,~{\rm and}~ \epsilon_k\rightarrow -\epsilon_k\quad{\rm for}~k=5,9.
\label{splitfundmentals}
\fe
We denote the invariant subset by $s_1$ and its complement by $s_2$. The local contribution from the irregular twisted singularity to the Coulomb branch dimension is
\ie
\dim_{\tilde\rho} {\rm Coulomb}=&\sum_{i\in s_1} (d_i-1)\ell+\sum_{i\in s_{2}}  \left( (d_i-1)\ell+1/2 \right)
\\
=&{  \ell h( J )\rank (J)\over 2}+{|s_2|\over 2}
\label{twistedCBdim}
\fe
where $|s_2|$ is the size of the set $s_2$ and the $i$-th summand gives the order of the highest pole in the differential $\epsilon_i$ whose coefficient is not purely determined by the polar matrices $T_m$ and $U_n$ from $\tilde{\rho}$ \cite{Witten:2007td}. 

As for degenerations of the maximal twisted irregular singularity, we have the following \textit{conjectured} formula, in analogy to the untwisted case, for counting the local contribution to the Coulomb branch dimension in terms of semi-simple orbits of $T_i$ in $\mfg^\vee$ \footnote{
	Heuristically the RHS of \eqref{Z2CBfromOrbits} counts the generalized monodromy data \cite{Witten:2007td} in the presence of the $\mZ_2$ twist. The appearances of $|s_2|$ which counts the $\mZ_2$ odd Cartan elements of $J$ can be understood as follows: as components of $U_j$ with $2\leq j\leq \ell$, they are parameters of the singularity rather than moduli.},
 \begin{tcolorbox}
	\ie\label{Z2CBfromOrbits}
	 &\dim_{\tilde\rho} {\rm Coulomb}
	 =
 {1\over 2}\Bigg(\sum_{i=1}^\ell  
	\dim  T_i +\sum_{j=2}^\ell\left(\dim J/\mfg^\vee-|s_2|\right)  
	+\dim J/\mfg^\vee 
	\Bigg) 
	\fe
\end{tcolorbox}
 
\noindent 
For $T_i$ regular semisimple, we have $\dim T_m=(\dim \mfg^\vee-\rank \mfg^\vee)/2$ and the above reduces to \eqref{twistedCBdim}.

Therefore the total Coulomb branch dimension of the AD theory, constructed from a maximal twisted irregular singularity and a principal regular singularity is
\ie\label{maxCBtot}
\dim   {\rm Coulomb}
=&{  (\ell+1) h( J )\rank(J)\over 2}+ |s_2|  
-\dim J
\\
= & {\rank(J)((\ell-1) h( J )-2)\over 2}+ |s_2|  
.
\fe 
On the other hand, the total Higgs branch quaternionic dimension of this AD theory is given by 
\ie\label{twistedirregHiggs}
\dim  {\rm Higgs}=&{1\over 2} (\dim \mfg-\rank \mfg)+\rank \mfg^\vee
\fe

The detailed Coulomb branch spectrum can be obtained from the SW curve as before. Below we take $ J =D_n$ for illustration. The singular SW curve, up to transformations that fix $xdz$ takes the form \footnote{
	Note that unlike the case with a single untwisted irregular singularity, the singular AD curve with all deformations turned off can not specify the full theory. However, one may still use the singular curve to read off the scaling dimensions of $x$ and $z$.
}
\ie\label{DmaxtwistedSW}
x^{2n} +x^2 z^{2(n-1)(\ell-2) } =0
\fe
which fixes the scaling dimensions
\ie{}
[x]={\ell-2\over \ell-1},~~[z]={1\over \ell-1}.
\fe
The crucial difference from the untwisted cases in the previous sections is that among the deformations of the singular SW curve, the Pfaffian $\tilde{\epsilon}_n$ is constrained to have half integer powers of $z$. With this in mind, we can quickly enumerate the Coulomb branch operators, in particular the total number of them is, from $\epsilon_{2i}$ and $\tilde{\epsilon}_n$
\ie
\dim {\rm Coulomb}=&\sum_{i=1}^{n-1}(2i(\ell-1)-\ell)+(n-1)(\ell-1)
\\
=&(n-1)(n(\ell-1)-1)
\fe
which agrees with \eqref{maxCBtot}. Furthermore, we have $2n-2$ mass parameters: $n-1$ of them correspond to the Casimirs of $USP(2n-2)$ flavor symmetry and the other $n-1$ of them come from the irregular singularity. From {\eqref{twistedirregHiggs}, the total Higgs branch quaternionic dimension is
	\ie 
	\dim  {\rm Higgs}= {1\over 2} (\dim C_{n-1}-\rank C_{n-1})+ n-1 =n(n-1)
	\fe
	
	We discuss some details about the $\mZ_2$ twisted regular singularities and the resulting AD theories in Appendix~\ref{app:exptwist}. We leave the generalizations to the $\mZ_2$ twist for $A_{2n}$ theories and the $\mZ_3$ twist for $D_4$ theories to the interested readers.
	
	\subsection{$D$ type twisted irregular singularities}
	The generalization to cases with leading polar matrix \textit{nilpotent} is straightforward for $D$ type theories since the $\mZ_2$ outer-automorphism of $SO(2n)$ can be identified with $O(2n)/SO(2n)$. In other words, the condition \eqref{condontwisted} becomes
	\ie
	\Phi(ze^{2\pi i})=\tilde g\Phi(z)\tilde g ^{-1},\quad 
	\fe
	with $\tilde g\in O(2n)$ and $\det \tilde g=-1$
	
	The twisted version of $D_n$ singularity with slope denominator $b=n$, takes the following form  with $k$ odd \footnote{
		The singular boundary condition for Higgs field with $k$ even must be accompanied by a gauge transformation $\tilde g\in SO(2n)$ which leads to the untwisted irregular singularity considered before.}, 
	\ie
	&\Phi={1\over z^{2+{k\over 2n}}}\begin{pmatrix}
		0& 1 & & & & & &  \\
		-1& 0 & & & & & &  \\
		&   & 0 & \omega^k & & & &  \\
		&   & -\omega^k & 0 & & & &  \\
		&   &   &   & & \ddots & &  \\
		&   &   &   & &  & 0 & \omega^{k(n-1)}  \\
		&   &   &   & &  & -\omega^{k(n-1)} & 0  \\
		
	\end{pmatrix}+\dots,
	\\
	&\omega^{2n}=1 
	\label{twistedDb=n}
	\fe
	with $O(2n)$ gauge transformation
	\ie
	\tilde g=\left(
	\begin{array}{c | c}
		0 & I_{2n-2} \\ \hline
		J_2  & 0
	\end{array}\right)
	\fe
	where we defined $I_{m}$ as the $m\times m$ identity matrix and $J_2=\begin{pmatrix} 1 & 0\\ 0 & -1
	\end{pmatrix}$.

	Alternatively we can also twist the irregular singularities with $b=2n-2$, and obtain for arbitrary $k\in \mZ$,
	\ie
	&\Phi={1\over z^{2+{k\over 2n-2}}}\begin{pmatrix}
		0& 0 & & & & & &  \\
		0& 0 & & & & & &  \\
		&   & 0 & 1 & & & &  \\
		&   & -1 & 0 & & & &  \\
		&   &   &   & & \ddots & &  \\
		&   &   &   & &  & 0 & \omega^{k(n-2)}  \\
		&   &   &   & &  & -\omega^{k(n-2)} & 0  \\
		
	\end{pmatrix},
	\\
	&\omega^{2n-2}=1
	\label{twistedDb=2n-2}
	\fe
	with $O(2n)$ gauge transformation 
	\ie
	\tilde g=\left(
	\begin{array}{c | c | c}
		I_2 & 0 & 0 \\ \hline
		0 & 0 &  I_{2n-4}  \\ \hline
		0 & J_2 & 0 \\  
	\end{array}\right)
	\fe
	for $k$ odd and
	\ie
	\tilde g=\left(
	\begin{array}{c | c | c}
		J_2 & 0 & 0 \\ \hline
		0 & 0 &  I_{2n-4}  \\ \hline
		0 & I_2 & 0 \\  
	\end{array}\right)
	\fe
	for $k$ even.

	For example consider the AD theory constructed from a twisted $D_n$ singularity of the form \eqref{twistedDb=2n-2} with $k=1$ and a simple regular puncture whose pole structure is $\{1,\dots,1,1/2\}$. The singular SW curve is
	\ie
	x^{2n}+x^2 z=0
	\fe 
	which fixes 
	\ie
	&[x]={1\over 2n-1},\quad [z]={2n-2\over 2n-1}.
	\fe
	The AD theory has $n-1$ Coulomb branch operators with dimensions $2i/(2n-1)$ with $n\leq i\leq 2n-2$ and one mass parameter \footnote{The Coulomb branch spectrum of this twisted theory is identical to that of the $(A_1,D_{2n-1})$ theory but we believe they have different BPS charge lattices thus are different $\cN=2$ SCFTs.}.
	
	Similarly, we can start with a twisted $D_n$ singularity of the form \eqref{twistedDb=n} with $k=1$ and a simple regular puncture. The singular SW curve is \footnote{
		Let us emphasize again that the ``singular'' SW curve here is only used to fix the scaling dimensions of $x$ and $z$.}
	\ie
	x^{2n}+  z=0
	\fe 
	which fixes 
	\ie
	&[x]={1\over 2n+1},\quad [z]={2n \over 2n+1}.
	\fe
	The AD theory has $n-1$ Coulomb branch operators with dimensions $2i/(2n+1)$ with $n+1\leq i\leq 2n-1$ and no mass parameters \footnote{As in the previous example, the Coulomb branch spectrum of this twisted theory is identical to that of the $(A_1,A_{2n-2})$ theory but we believe they have different BPS charge lattices thus are different $\cN=2$ SCFTs.}.
	%
	
	It is also straightforward to construct AD theories from the above twisted irregular singularities in the presence of twisted full regular singular singularities. See Figure~\ref{fig:twisted4$d$4theories} for examples of the Newton polygons for these theories. Note that in contrast to those of the untwisted theories in Figure~\ref{fig:type4$d$4theories}, the $x$-independent monomials that would contribute to the SW curve now correspond to half-filled dots that have been shifted below by one unit because the Pfaffian invariant $\tilde\epsilon_n$ is odd under a $\mZ_2$ twist.

	\begin{figure}[htb]
		\centering
		{
			\begin{minipage}{0.45\textwidth}
				\centering
				\includegraphics[scale=.8 ]{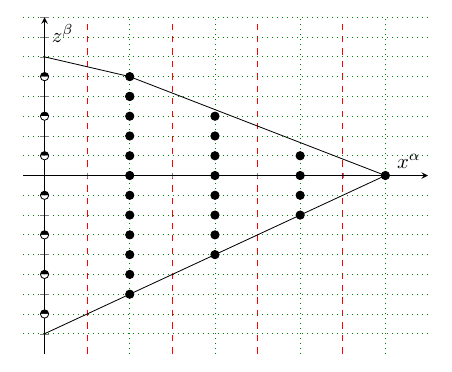}\\
			\end{minipage} 
		}
		\centering
		{
			\begin{minipage}{0.45\textwidth}
				\centering
				\includegraphics[scale=.8 ]{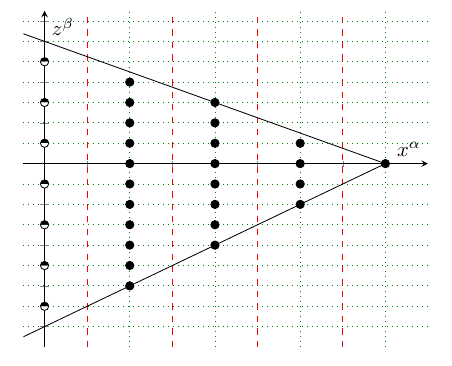}\\
			\end{minipage} 
		}
		\caption{Twisted $D_4$ theories with twisted full regular singularities.}
		\label{fig:twisted4$d$4theories}
	\end{figure}

	\section{More Properties of AD Theories}\label{sect:V}
	\subsection{Coulomb branch spectrum}
	As we have discussed in the previous sections, for $D$ type theory it is straightforward to read off the spectrum of the AD theory from an irregular singularity and possibly an additional regular singularity using the Newton polygon. For $E$ type theories, the spectral curve representation of the spectrum is rather redundant, in which case it is more convenient to use the IIB 3-fold  singularity description whose complex structure deformations constitute the full Coulomb branch spectrum with no redundancy \footnote{
		In the spectral curve, we need to remove redundant deformations due to coordinate redefinitions that leave $\lambda=xdz$ invariant and also trace relations among the invariant differentials.
	}. The scaling dimensions of the Coulomb branch parameters which correspond to \textit{certain} coefficients in $\epsilon_i(z)$ can be read off easily following the procedure in subsection~\ref{IIBcon}. However, some of the deformation parameters have negative 4$d$ scaling dimension and correspond to irrelevant couplings which must be removed from the list of physical Coulomb branch parameters.
	
	To incorporate an additional regular singularity, we can simply allow $\epsilon_i(z)$ to have a pole in $z$ according to the pole structure $\{p_{d_i}\}$ with $1\leq i \leq \rank J$ associated to the regular singularity, 
	\ie
	\epsilon_i(z)=\dots+{u_{i,1}\over z}+{u_{i,2}\over z^2}\dots+{u_{i,p_{d_i}}\over z^{p_{d_i}}}.
	\fe
	The $u_{i,j}$'s are unconstrained Coulomb branch parameters whose scaling dimensions can again be easily fixed (all positive). Together with the parameters associated with the isolated 3-fold singularity in the absence of poles in $\epsilon_i(z)$, they make up the entire Coulomb branch spectrum.

	\subsection{Central charges}
	There are a number of ways to compute the central charges of the AD theories that we have constructed. Some of them are more useful than others depending on the input.
	
	When there exists a weak coupling description, the central charges $a,c$ are determined by \cite{Shapere:2008zf}
	\ie
	2a-c={1\over 4}\sum_{i}(2[u_i]-1),~~~a-c={1\over 24}(n_v-n_h)
	\label{weakcouplingac}
	\fe
	where $[u_i]$ denotes the scaling dimension of the Coulomb branch operator $u_i$, $n_v$ counts the number of vector multiplets and $n_h$ counts the number of hypermultiplets. More generally, when the theory has a Higgs branch and is completely Higgsed, we can rewrite the second equation as \cite{Xie:2012hs,Xie:2013jc}
	\ie
	a-c=-{\dim_\mH{\rm Higgs} \over 24}.
	\label{aminuscHiggs}
	\fe 
	For AD theories constructed from Type $J$ (2,0) SCFTs on $\mP^1$ with an integral pole at the irregular singularity, we expect $\dim_\mH{\rm Higgs}=\rank J$.
	
	For strongly coupled theories, there is a formula for the central charges from topological field theories \cite{Shapere:2008zf},
	\ie
	a={R(A)\over 4}+{R(B)\over 6}+{5r\over 24}+{h\over 24},~~~c={R(B)\over 3}+{r\over 6}+{h\over 12}
	\label{strongcouplingac}
	\fe
	where $R(A),R(B)$ are the $R$-charges of path integral measure factors and $r,h$ are the number of free vector-multiplets and hypermultiplets at {\it generic} points of the Coulomb branch. For the theories we consider, $r$ coincides with the rank of the Coulomb branch and $h$ is 0. Moreover $R(A)$ can be expressed in terms of the scaling dimensions of the Coulomb branch operators,
	\ie
	R(A)=\sum_i ([u_i]-1)
	\fe
	For generic strongly coupled $\cN=2$ SCFTs, it is difficult to compute $R(B)$. However we have the following formula for $(G,G')$ theories from \cite{Cecotti:2013lda},
	\ie
	R(B)={1\over 4}{r(G)r(G')h(G)h(G')\over h(G)+h(G')}
	\fe
	
	This formula has an elegant extension for general isolated hypersurface singularities in the IIB description \cite{danyau}. Given a IIB singular 3-fold defined by $W(x_1,x_2,x_3,x_4)=0$ in $\mC^3$ which has a $\mC^*$ action with positive charges $\{q_i\}$, we can compute $R(B)$ by \cite{danyau}:
	\ie
	R(B)={\m \A\over 4}
	\label{RBfromIIB}
	\fe
	where $\m$ is the Milnor number for $W$ and 
	\ie
	\A={1\over \sum_{i=4}^r q_i-1}
	\fe
	is the scaling dimension of the constant deformation. 
	
	Since all of the AD theories from untwisted irregular singularity considered here have IIB description in terms of isolated hypersurface singularities, we can extract their $a$ and $c$ central charges from \eqref{RBfromIIB} and \eqref{strongcouplingac}.
	
	As for the twisted theories, generally only the $2a-c$ anomaly can be obtained from the Coulomb branch spectrum and more techniques need to be developed in order to compute $a$ and $c$ separately, for example, 3d mirror symmetry for the $S^1$ reduction of these AD theories would be a useful tool. Below we are focus on the central charges for the untwisted theories (see Appendix~\ref{app:exptwist} for examples in the twisted case).

	It is known for cDV singularities of index $n$ (i.e. $cA_n,cD_n,cE_n$) with the additional coordinate $z$, that its Milnor number is given by \cite{arnold2012singularities}
	\ie
	\m=n\left({1\over q(z)}-1\right)
	\fe
	where $q(z)$ is the $\mC^*$ charge of $z$. For example, for $b=h$ the Coxeter number, the Milnor number is simply
	\ie
	\m=n(k-1).
	\fe
	In general we have $q(z)={b\over h k}$ and $\A={hk\over b+k}$, thus
	\ie
	\m=n\left( {hk\over b}-1\right),
	\fe
	which leads to
	\ie
	R(B)={nhk(hk-b)\over 4b(b+k)}
	\fe
	using \eqref{RBfromIIB}.
	
	\begin{table}[htb]
		\begin{center}
			\begin{tabular}{ |c|c| c|c|  }
				\hline
				$J  $ & $\m(b_1)$ & $\m(b_2)$ & $\m(b_3)$
				\\ \hline
				$ A_n  $ & $n(k-1)$  & $(n+1)k-n$  &    \cellcolor{gray!50}    \\ \hline
				$D_n$ & $n(k-1)$  & $(2n-2)k-n$ & \cellcolor{gray!50}  \\ \hline
				$E_6$  &  $6(k-1)$   & $8k-6$ &  $9k-6$ \\ \hline
				$E_7$ &  $7(k-1)$ & $9k-7$ & \cellcolor{gray!50}  \\ \hline
				$E_8$  &  $8(k-1)$  & $10k-8$  & $12k-8$ \\ \hline
			\end{tabular}
		\end{center}
		\caption{Milnor numbers (BPS lattice dimension) for $J^{(b)}[k]$ theories.}
		\label{milnor}
	\end{table}

	\subsubsection{Examples from the maximal irregular singularities}\label{sect:featMAX}
	In general it is straightforward to compute $R(A)$ using the Coulomb branch spectrum obtained from either the spectral curve or the associated 3-fold singularity. Although a closed form expression of $R(A)$ for general $J^{(b)}[k]$ is not available at present, for the special subclass of theories $J^{(b)}[bm]\cong (J,A_{hm-1})$ with $m\in\mZ^+$ which originate from maximal irregular singularities introduced in section~\ref{sect:maximalut}, the problem is vastly simplified.
	
	Since $J^{(b)}[bm]$ does not depend on the (allowed) choice of $b$ up to marginal deformations, its Coulomb branch spectrum is captured uniformly by a single spectral curve or its corresponding 3-fold cDV singularity for given $J$ and positive integer $m$. In particular, following the procedure outlined in the previous section, it is easy to see that the Coulomb branch of $J^{(b)}[bm]$ has dimension $n(mh/2-1)$ among which there are $n-1$ marginal operators\footnote{
		The case $A_n^{(b)}[b]$ is an exception, which has $n-2$ marginal operators on the Coulomb branch.} and $mn-1$ relevant operators (see Table~\ref{tab:maxSPEC}). Moreover, from the Coulomb branch spectrum one can compute from \eqref{strongcouplingac}, the $a$ and $c$ central charges for $J^{(b)}[bm]$ theory with $J$ of rank $n$,
	\ie
	c=\frac{n \left(h (h+1) m^2-2 m-2\right)}{12 (m+1)}
	,\quad a=\frac{n \left(2 h (h+1) m^2-3 m-3\right)}{24 (m+1)}
	\fe
	which satisfies \eqref{aminuscHiggs} with $\dim_\mH{\rm Higgs} =\rank (J)$ in agreement with our expectation.
	\begin{table}[htb]
		\begin{center}
			\begin{tabular}{ |c|c| c|c|c|c|c|c|c| }
				\hline
				$J$ &   $r$ & $r_{marg}$ &  $r_{rel}$ & $n_f$  
				\\ \hline
				$ A_n  $ & ${1\over 2}n(m(n+1)-2)$ & \txt{$n-1$ ($m>1$) \\$n-2$ ($m=1$)}&  $mn-1$ & $n$     \\ \hline
				$D_{n>3}$ & $n(m(n-1)-1)$ & $n-1$  &  $mn-1$  & $n$  \\ \hline
				$E_6$  &  $6(6m-1)$  & 5 &  $6m-1$  & $6$   \\ \hline
				$E_7$ &  $7(9m-1)$ & 6 & $7m-1$ & $7$   \\ \hline
				$E_8$  &  $8(15m-1)$ & 7 &  $8m-1$ & $8$   \\ \hline
			\end{tabular}
		\end{center}
		\caption{Coulomb branch dimension $r$, number of marginal operators $r_{marg}$ and relevant operators $r_{rel}$ for $J^{(b)}[bm]$ theories.}
		\label{tab:maxSPEC}
	\end{table}

	\subsubsection{Limits of the central charges}
	Here we will consider various large parameter limits of the $J^{(b)}[k]$ theories and obtain the asymptotic behaviors for central charges $a$ and $c$. In particular we see that $a=c$ in these limits.
	
	Let us start with the limit $k\rightarrow \infty$ with $n$ being finite. In this limit we have 
	\ie
	\A={hk\over b+k}\sim h,\quad r\sim {\m\over 2}={nhk\over 2b}
	\fe
	which gives
	\ie
	R(B)={\m \A \over 4}\sim {nh^2k\over 4b }.
	\fe
	The Coulomb branch spectrum (in this limit $[x]\sim 1,[z]\sim b/k  $) from the spectral curve is given by
	\ie
	\{d_i-{bj\over k}| 1\leq i\leq n,~j\geq 1~{\rm such~that}~d_i-{bj\over k}>1\}
	\label{largekspec}
	\fe
	where $d_i$'s are the degrees of fundamental invariants in Table~\ref{liedata}. 
	From \eqref{largekspec} we can derive
	\ie
	R(A)=&\sum_\A ([u_\A]-1)
	\\
	\sim& \sum_{i=1}^n \left(
	(d_i-1) (b_i+1)-{b_i(b_i+1)\over 2}{b\over k}
	\right)
	\fe
	where $b_i$ counts the number of Coulomb branch operators from the invariant differential $\epsilon_i$. In the limit $k\rightarrow \infty$, we have
	\ie
	b_i\sim (d_i-1){k\over b}
	\fe
	which implies
	\ie
	R(A)\sim {k\over 2b}\sum_{i=1}^n  (d_i-1)^2
	={k  n h(2h-1)\over 12b} 
	\fe
	where we have used the Lie algebra identities
	\ie
	\sum_{i=1}^n d_i={1\over 2}n(h+2),\quad \sum_{i=1}^n d_i^2={1\over 6}(2h^2+5h+6)n.
	\fe
	Therefore the central charges $a$ and $c$ are determined to be
	\ie
	a=c={knh(h+1)\over 12 b}.
	\fe
	In particular, for the maximal slope denominator $b=h$, we have
	\ie
	a=c={kn (h+1)\over 12  }.
	\fe
%
	Next let us inspect the $n\rightarrow \infty$ limit with $k$ being finite in the $A$ and $D$ type theories.
	For $D_n^{2n-2}(k)$ and $D_n^{n}(k)$ theories, by studying the deformations of the spectral curves, we have in this limit
	\ie
	&{\rm for}~ D_n^{(2n-2)}[k]_{n\rightarrow \infty} \quad a=c={(k^2-1)n\over 12} ,
	\\
	&{\rm for}~ D_n^{(n)}[k]_{n\rightarrow \infty} \quad~~~~ a=c={(4k^2-1)n\over 12}.
	\fe
	Similarly for $A_n^{n+1}(k)$ and $A_n^{n}(k)$ theories, we have\footnote{
		Note that the two theories $D^{(2n-2)}_n[2k]$ and $D^{(n)}_n[k]$ are \textit{approximately} equivalent in the limit $k\rightarrow \infty$. This is most easily seen from the associated c$D_n$ singularities. The same statement applies to $A_n^{(n+1)}[k]$ and $A_n^{(n)}[k]$.}
	\ie
	&{\rm for}~ A_n^{(n+1)}[k]_{n\rightarrow \infty} \quad a=c={(k^2-1)n\over 12} ,
	\\
	&{\rm for}~ A_n^{(n)}[k]_{n\rightarrow \infty} \quad~~ ~a=c={(k^2-1)n\over 12} .
	\fe

	Finally we consider theories $A_n^{(b)}[bm]$ and  $D_n^{(b)}[bm]$ in the $n\rightarrow \infty$ limit (thus $b\rightarrow\infty$) with $m$ being finite (which corresponds to taking $k$ and $\rank(J)$ large with their ratio fixed). The central charges in this limit can be easily read off from Table~\ref{tab:maxCC}: 
	\ie
	&{\rm for}~ A_n^{(b)}[bm]_{n\rightarrow \infty} \quad a=c={m^2 n^3\over 12(m+1)} ,
	\\
	&{\rm for}~ D_n^{(b)}[bm]_{n\rightarrow \infty} \quad a=c={m^2 n^3\over 3(m+1)}.
	\fe

	\section{Conclusion and Discussions}\label{sect:VI}
	Using M5 branes, we have constructed a large class of new $\cN=2$ SCFTs by classifying 
	the irregular punctures. We have also given the corresponding 3-fold hypersurface singularities in the IIB description. 
	Along the way, we have established a map between the irregular singularities of Hitchin system, Argyres-Douglas theories, and isolated 
	hypersurface singularities (see Figure~\ref{fig:ADsh}). 
	The main purpose of this paper is to  give a classification of the possible theories within this construction, and there are many other 
	interesting questions about these theories that one can study.
	
	Some of the theories (e.g. from maximal irregular singularities) constructed here have exact marginal deformations, and one question is to identify the corresponding 
	duality group. It is expected that one can find many weakly coupled 
	gauge theory descriptions, and it is interesting to study them systematically (see \cite{Buican:2014hfa} for some examples). 
	
	A special subclass of our theories, labeled by $(J^{(b)}[k],F)$ with $k=-b+1$, are rigid matters in the sense that they do not contain any Coulomb branch moduli but have full flavor symmetry $J$.\footnote{
		This was the rigid local model studied in \cite{2009arXiv0901.2163F}.} By gauging a diagonal subgroup of the flavor symmetries of two such rigid matter systems, one may generate $\cN=2$ asymptotically free gauge theories. With outer-automorphism twist, we can construct $\cN=2$ asymptotically free gauge theories for arbitrary gauge group $G$ this way.\footnote{
		The case with $b=h(J)$ gives rise to pure SYM, whereas the other values of $b$ lead to SYM coupled to matter.}
	
	The RG flows between the AD theories have been explored in \cite{Xie:2013jc} for $A$ type theories by considering various deformations of the singular SW curve. In terms of the IIB isolated 3-fold singularities, these flows are captured by the so-called adjacency relations\footnote{
		It can be thought of as an inclusion relation between the deformation space of the singularities.} between different singular varieties. In particular, some of the adjacency relations among Arnold's simple singularities (which correspond to cDV singularities labeled by $J^{(h)}[2]$ ) were realized explicitly by RG flows in \cite{Xie:2013jc}. We expect a similar relation between adjacency relations among the 3-fold singularities considered in this paper (Table~\ref{table:isolatedsingularitiesALEfib}) and the RG flows among the corresponding AD theories.
	
	It is interesting to study various partition functions of these theories, and we expect that
	our M5 brane construction would be quite useful. In particular, we expect that the 
	two point function (with insertions of operators corresponding to our irregular punctures) of the 2d Toda theory would give 
	the $S^4$ partition function \cite{Alday:2009aq}. Similarly, the two point function of the $q$-deformed YM theory would give 
	the superconformal (Schur) index \cite{Gadde:2011ik} (see \cite{Buican:2015ina,Cordova:2015nma} for recent result on $A_1$ type AD theory). 
	Once the index is obtained, it is interesting to find the corresponding chiral algebra \cite{Buican:2015ina,Cordova:2015nma}. 
	
	For $A$ type theory whose irregular singularity has integer order poles, one can compactify the theory 
	on a circle to get a 3$d$ $\cN=4$ SCFT. The mirror for these theories has been written down in \cite{2007arXiv0706.2634B,Benini:2010uu,Xie:2012hs} and 
	they are all Lagrangian quiver gauge theories. We expect that the new theories engineered here using integer order pole irregular 
	singularity (the maximal irregular singularities and their degenerations) will also have three dimensional mirrors, and it would be interesting to develop a systematic identification. 
	
	For our theories $J^{(b)}[k]$, their central charges satisfy the condition $a=c$ in the large $k$ or large $\rank(J)$ limit, and this indicates
	that the theories may have supergravity duals \cite{Gaiotto:2009gz}. It would be very interesting to derive the supergravity dual explicitly.  
	
	As we have briefly mentioned, for the same three-fold isolated quasi-homogeneous singularity, had we kept the string scale finite while decoupling gravity, we would end up with a 4$d$ non-gravitational string theory (known as little string theory or LST for short) whose low energy limit gives the 4$d$ $\cN=2$ SCFT \cite{Ooguri:1995wj,Giveon:1999zm,Kutasov:2001uf,Hori:2002cd}. There we have another holographic picture in terms of type II string theory on linear dilaton background with an $\cN=2$ LG sector which is well-defined even at finite $k$ and $\rank(J)$. It would be very interesting to understand what kind of dynamics in the 4$d$ $\cN=2$ AD theory (as a low energy sector of the full LST)  we can learn from the bulk string theory description, possibly in a double-scaled limit (to cap off the dilaton throat) along the lines of \cite{Giveon:1999px,Aharony:2003vk,Aharony:2004xn,Chang:2014jta,Lin:2015zea}.
	
	Finally, the irregular singularities of the Hitchin system considered here are codimension-two half-BPS defects of the 6$d$ $(2,0)$ theory. Upon compactification on $T^2$ longitudinal to the defect, we obtain a half-BPS surface operator in 4$d$ $\cN=4$ SYM \cite{Chacaltana:2012zy}. It would be interesting to study the surface operators obtained this way from our irregular singularities, especially the ones with outer-automorphism twist, in relation to the geometric Langlands program \cite{Witten:2007td,2009arXiv0901.2163F}.

	\section*{Acknowledgments}
	We would like to thank Bruno Le Floch for helpful comments on the first version of this work. We are grateful to the Weizmann Institute of Science, the 32nd Jerusalem Winter School in Theoretical Physics, and the Simons Summer Workshop in Mathematics and Physics 2015 for the hospitality during the course of this work. We also would like to thank  the Center for the Fundamental Laws of Nature at Harvard.
	Y.W. is supported in part by the U.S. Department of Energy under grant Contract Number  DE-SC00012567. The work of DX is supported by the Center for Mathematical Sciences and Applications at Harvard University.

	\appendix
	
	\section{Isolated Quasi-homogeneous cDV Singularties}\label{apx:isocdv}
	In this section, we prove that the isolated quasi-homogeneous cDV singularities defined by
	\ie
	W_J(x_1,x_2,x_3,z)=f_J(x_1,x_2,x_3)+zg(x_1,x_2,x_3,z)
	\label{cDVapp}
	\fe
	with $J=A,D,E$,
	are precisely those listed in Table~\ref{table:isoscDV} (up to marginal deformations).
	
	First, it is easy to check that the quasi-homogeneous singularities in Table~\ref{table:isoscDV} all have finite Milnor numbers thus isolated.
	
	A necessary condition for a general quasi-homogeneous singularity $W(x_i)=0$ to be isolated is that for any axis $x_i$ there must be at least one monomial $\prod_j x_j^{k_j}$ in $W(x_i)$ such that $\sum_j k_j- k_i\leq 1$, otherwise there will be a singular locus along the $x_i$ axis \cite{arnold2012singularities}. 
	
	Now given a quasi-homogeneous cDV singularity defined by $W_J$ in \eqref{cDVapp} (in particular $q(x_i),q(z)>0$), $W_J(x_i,z)$ must contain monomial(s) from the set $L=\{z^k, z^k x_1,z^k x_2,z^k x_3\}$ with $k\geq 1$ to avoid a singular locus along the $z$ axis.  
	
	\begin{enumerate}
		\item For $W_J(x_i,z)$ of c$A_n$, c$E_6$ and c$E_8$ types, up to coordinate redefinitions, such $W_J(x_i,z)$ is always captured by the normal forms (or their marginal deformations) in Table~\ref{table:isoscDV}
		\item For $W_J(x_i,z)$ of c$D_n$ type, if $W_J(x_i,z)$ contains any of the three monomials $z^k,z^k x_1,z^k x_3$, up to a coordinate transformation, such $W_J(x_i,z)$  is captured by the two normal forms (or their marginal deformations) in Table~\ref{table:isoscDV}.
		However if $W_J(x_i,z)$ only contains the $z^k x_2$ monomial from the set $L$, then up to a coordinate transformation, we may assume that the $z$ dependent monomials of $W_J(x_i,z)$ are all of the form $z^i x_2^j$ with $i,j\geq 1$. Consequently, we have a singular locus along $x_3^2+z^k=x_2=x_1=0$. 
		
		\item For $W_J(x_i,z)$ of c$E_7$ type, if $W_J(x_i,z)$ contains any of the three monomials $z^k,z^k x_1,z^k x_3$, up to a coordinate transformation, such $W_J(x_i,z)$  is captured by the two normal forms (or their marginal deformations) in Table~\ref{table:isoscDV}.
		However if $W_J(x_i,z)$ only contains the $z^k x_2$ monomial from the set $L$, then up to a coordinate transformation, we may assume that the $z$ dependent monomials of $W_J(x_i,z)$ are all of the form $z^i x_2^j$, $z^i x_2^j x_3$ or $z^i x_3^2$ with $i,j\geq 1$. If $W_J(x_i,z)$ does not contain monomials of the form $z^i x_3^2$, we again end up with a singular locus along $x_3^3+z^k=x_2=x_1=0$; otherwise the $\mC^*$ charges $q(x_2)=1/3$ and $q(x_3)=2/9$ demand a term of the form $z^{6m}x_2+z^{5m}x_3^2$ in $W_J(x_i,z)$ for some $m\in\mZ^+$ (i.e. $k\in6\mZ$), in which case $W_J(x_i,z)$ is simply a marginal deformation of $f_{E_7}(x_i,z)+z^{9m}$.
	\end{enumerate} 
	
	Hence we have completed the proof.
	
	%
	

	\section{Examples of AD Theories from Maximal Twisted Irregular Singularities}\label{app:exptwist}
	For the AD theory engineered using a $D$ type maximal twisted singularity and another full regular twisted singularity considered in section~\ref{sect:maxtwist} with singular SW curve \eqref{DmaxtwistedSW},
	the dimensions of the Coulomb branch operators are,
	\ie
	&\{2i-{k\over \ell-1}| 1\leq i\leq n-1,~k\geq 1~
	\\
	&{\rm such~that}~2i-{k\over \ell-1}>1\}
	\fe
	from $\epsilon_{2i}$ and
	\ie
	\{n-{2k+1\over 2(\ell-1)}|  k\geq 0~{\rm such~that}~n-{2k+1\over 2(\ell-1)}>1\}
	\fe
	from $\tilde{\epsilon}_n$. Hence we have from \eqref{weakcouplingac},
	\ie
	2a-c=&{1\over 4}\sum_j (2[u_j]-1)
	\\
	=&\frac{1}{12} (n-1) n (4 (\ell-1) n-2 \ell-1)
	\fe
	and from \eqref{twistedirregHiggs} and \eqref{aminuscHiggs}
	\ie
	a-c=&-{n(n-1)\over 24},
	\fe
	allowing us to determine $a$ and $c$ for this class of AD theories
	\ie a=&\frac{1}{24} (n-1) n (8 (\ell-1) n-4 \ell-1),\\ 
 	c=& \frac{1}{6} (n-1) n (2 (\ell-1) n-\ell).
	\fe
		It is straightforward to repeat the above analysis for twisted $A_{2n-1}$ and $E_6$ theories from an irregular twisted singularity with regular semisimple polar matrices $T_i$ (in $\mfg^\vee$) and a regular twisted puncture of general type. Suppose the pole structure associated with the regular twisted singularity is denoted by $\{p_{d_i}\}$ with $1\leq i\leq \rank (J)$, then the Coulomb branch spectrum is given by 
	\ie
	\{p_{d_i}+1-{k\over \ell-1}|d_i\in s_1,~k\geq 1~{\rm such~that}~p_{d_i}-{k\over \ell-1}>0\}
	\fe
	from the $\mZ_2$ invariant differentials and
	\ie
	&\{p_{d_i}+{1\over 2}-{2k+1\over 2(\ell-1)}|d_i\in s_2,~k\geq 0~
	\\
	&{\rm such~that}~p_{d_i}+{1\over 2}-{2k+1\over 2(\ell-1)}>1\}
	\fe
	from the $\mZ_2$ odd differentials. In addition, we have $2|s_1|=2(r-|s_2)$ mass parameters, half of which correspond to the Casimirs of either $SO(2n+1)$ or $F_4$ flavor symmetry (the other half are associated with the twisted irregular singularity). The Higgs branch quaternionic dimensions are conjectured to be $n(n+1)$ for the twisted $A_{2n-1}$ theory and $(52-4)/2+4=28$ for twisted $E_6$ theory.

  \bibliography{ADrefs} 
 \bibliographystyle{JHEP}
\end{document}